\documentclass[a4paper,12pt]{article}
\usepackage{amsmath,amssymb} \usepackage{euler,eucal}
\usepackage[utf8]{inputenc}
\usepackage[T1]{fontenc}
\usepackage[english]{babel}
\usepackage{tgpagella}
\usepackage[super]{nth}
\usepackage[en-GB,showseconds=false,showzone=false]{datetime2} 
\DTMlangsetup[en-GB]{ord=raise}
\usepackage[left=20mm,right=20mm,top=20mm,bottom=20mm,a4paper]{geometry}
\usepackage[unicode=true,plainpages=false]{hyperref}
\usepackage{graphicx}
\hypersetup{colorlinks=true,pdfborder={0 0 0.1},linkcolor=orange!50!black,urlcolor=blue!50!black,citecolor=blue!50!black,
          pdftitle={Tensor product approach to modelling epidemics on networks},
          pdfauthor={S. V. Dolgov, D. V. Savostyanov},
}

\usepackage{tikz} \usetikzlibrary{arrows.meta,positioning,shapes,shadows,calc,decorations.text}
\usepackage{tikzpeople}
\pgfrealjobname{ttsir-arxiv1}

\def\Scol{blue}
\def\Icol{red}
\def\Rcol{green!50!black}
\tikzstyle{SSA}=[orange!50!black]
\tikzstyle{inf}=[thick,red!50!black]
\tikzstyle{rec}=[green!30!black]
\tikzstyle{S}=[fill=\Scol,draw=none,fill opacity=.3]
\tikzstyle{I}=[fill=\Icol,draw=none,fill opacity=.3]
\tikzstyle{R}=[fill=\Rcol,draw=none,fill opacity=.3]
\tikzstyle{Ss}=[inner color=\Scol!85!white,outer color=\Scol!15!white,draw=none,fill opacity=.65,shade]
\tikzstyle{Is}=[inner color=\Icol!85!white,outer color=\Icol!15!white,draw=none,fill opacity=.65,shade]
\tikzstyle{Rs}=[inner color=\Rcol!85!white,outer color=\Rcol!15!white,draw=none,fill opacity=.65,shade]
\tikzstyle{tick}=[scale=.8]
\tikzstyle{state}=[circle,minimum width=10mm,inner sep=.3em,
                   top color=white, bottom color=orange!30, 
                   text=black,text opacity=1,
                   draw=orange!50!black,thick,
                   circular drop shadow={opacity=.7}]
\tikzstyle{label}=[rectangle,rounded corners=2mm,minimum width=10mm,inner sep=.5em,
                   top color=white, bottom color=orange, fill opacity=.3,
                   text=black,text opacity=1,
                   draw=orange!50!black,thick]
\tikzstyle{lab}=[circle,top color=blue!60!black,bottom color=black,text=yellow,minimum size=2.2em,inner sep=0em,circular drop shadow={opacity=.7},scale=.6]
\tikzstyle{aim}=[inner sep=0mm,outer sep=.5mm]

\usepackage{pgfplots}
\pgfplotsset{%
      log number format basis/.code 2 args={${\pgfmathprintnumber{#2}}$},
      every axis/.append style={width=.475\textwidth,height=.35\textwidth,
                                axis x line=bottom, axis y line=left,
                                x axis line style={very thick,->}, y axis line style={very thick,->},
                                tick align=inside, tick style={thick},
                                every x tick label/.style={scale=.8},
                                every y tick label/.style={scale=.8},
                                major grid style={black,opacity=.5}, minor grid style={thin,opacity=.5},
                                grid=major,
                                cycle list name=ppve,
                                },
      every axis legend/.append style={
                                legend columns=1,
                                legend cell align={left},
                                draw=none,
                                fill=white,
                                },
      every axis x label/.style={at={(0.5,-.07)},below,fill=none,fill opacity=1,text opacity=1},
      every axis y label/.style={at={(-.1,1.07)},right,fill=none,fill opacity=1,text opacity=1},
      }
\pgfplotscreateplotcyclelist{ppve}{%
  thick, black,  opacity=1, \\ 
  ultra thick, black!70, opacity=.7, \\
  ultra thick,red,    opacity=.4, \\
  thick, red,         opacity=1,  \\
}
\def\hat{\mathaccent"705E }
\def\eps{\varepsilon}
\def\geq{\geqslant} \def\leq{\leqslant}
\def\Real{\mathbb{R}}
\def\X{\mathbb{X}}            
\def\O{\mathcal{O}}
   
\def\inf{\text{(inf)}} \def\rec{\text{(rec)}}   
\def\dt{\mathrm{d} t} 

\def\phi{\varphi}
\def\Prob{\mathsf{P}} \def\Mean#1{\mathsf{E}[#1]} \def\Var#1{\mathsf{V}[#1]} 
 \def\iff{\:\Leftrightarrow\:} 
\def\rank{\mathop{\mathrm{rank}}\nolimits}

\def\diag{\mathop{\mathrm{diag}}\nolimits}

\def\state#1{({\mathrm{#1}})} 
\def\evec{\vec e} \def\svec{\vec s} \def\ivec{\vec{\text{\emph{\i}}}}   
\def\x{\mathrm{x}} \def\y{\mathrm{y}} \def\e{\mathrm{e}}  
\def\p{\mathbf{p}} \def\q{\mathbf{q}} \def\a{\mathbf{a}} \def\z{\mathbf{z}} \def\f{\mathbf{f}} 
\def\J{\mathbf{J}} \def\A{\mathbf{A}} \def\Proj{\mathbf{P}} 
\def\Id{\mathrm{Id}}
\def\zero{\color{black!40}{0}}
\def\Indicator#1{\mathbf{1}_{#1}}
\def\Nodes{\mathcal{V}} \def\Edges{\mathcal{E}}  

\def\Nssa{N_{\mathrm{SSA}}}
\begin{document}
\author{Sergey V. Dolgov\footnotemark[2]~ and Dmitry V. Savostyanov\footnotemark[3]}
\title{Tensor product approach to \\ modelling epidemics on networks\thanks{%
        Equal contributions. The order of authors is alphabetical.
        SD was supported by the Engineering and Physical Sciences Research Council New Investigator Award EP/T031255/1.
        DS was supported by the Leverhulme Trust Research Fellowship RF-2021-258.
       }}
\date{\nth{30} August 2022}

\renewcommand{\thefootnote}{\fnsymbol{footnote}}
\footnotetext[2]{University of Bath, Claverton Down, Bath, BA2 7AY, United Kingdom (\href{mailto:s.dolgov@bath.ac.uk}{S.Dolgov@bath.ac.uk}).}
\footnotetext[3]{University of Essex, Wivenhoe Park, Colchester, CO4 3SQ, United Kingdom (\href{mailto:d.savostyanov@essex.ac.uk}{D.Savostyanov@essex.ac.uk}, \href{mailto:dmitry.savostyanov@gmail.com}{dmitry.savostyanov@gmail.com}).}
\renewcommand{\thefootnote}{\arabic{footnote}}

\maketitle

\begin{abstract}
 To improve mathematical models of epidemics it is essential to move beyond the traditional assumption of homogeneous well--mixed population and involve more precise information on the network of contacts and transport links by which a stochastic process of the epidemics spreads.
 In general, the number of states of the network grows exponentially with its size, and a master equation description suffers from the curse of dimensionality.
 Almost all methods widely used in practice are versions of the stochastic simulation algorithm (SSA),  which is notoriously known for its slow convergence.
In this paper we numerically solve the chemical master equation for an SIR model on a general network using recently proposed tensor product algorithms.
 In numerical experiments we show that tensor product algorithms converge much faster than SSA and deliver more accurate results, which becomes particularly important for uncovering the probabilities of rare events, e.g. for number of infected people to exceed a (high) threshold.

\end{abstract}


\section{Introduction}

Modelling of epidemics is crucial to inform policies and support decision making for disease prevention and control.
The recent outbreak of COVID-19 pandemic raised a significant scientific and public debate regarding the quality of the mathematical models used to predict the effect of the pandemics and to choose an appropriate response strategy.
One of the first epidemiological models, proposed by Kermack and McKendrick in 1927~\cite{kmk-sir-1927}, assumes that each member of the population can be either susceptible to a disease, infected, or recovered. 
Its second important assumption is that the population is \emph{well--mixed}, i.e. all members are in contact with each other and have the same chance of getting and passing a disease.
Under this assumption, the system dynamics is governed only by the sizes of the compartments for susceptible, infected, and recovered part of the population, and can be described by three ordinary differential equations, one for each compartment.
Despite its simplicity, the Kermack--McKendrick SIR model can describe important stages of the epidemics, such as exponential growth of the number of infected people at the beginning of epidemic, and the exponential decay after the epidemics passed its peak.
For this reason, this and other compartmental models are often included in academic curriculum and used to popularise epidemiological modelling among general public and present it to policy makers.

When it comes to policy making, however, we need models that can provide accurate quantitative results. 
The main assumption behind the compartmental models --- that the population is well--mixed --- does not hold very well for human population.
People are not in constant contact with each other, and the probability of two people to meet each other depends significantly on where they live, where they work and what social contacts they maintain.
The speed of the epidemics depends not just on the total number of infected people, but on where the infected people are located in relation to the susceptible part of the population. 
For example, if all infected people are located in an isolated region, the disease will spread much slower, than if the same number of infected people were spread uniformly among the susceptible part of the population.
Moreover, the contacts are inherently stochastic.
Due to this \emph{internal} noise, deterministic models can give wrong results for extreme scenarios of e.g. small number of rarely contacting infected individuals.

To deliver more accurate predictions,
\emph{stochastic epidemiological models on networks}
have been proposed \cite{chen-stoch-sir-2005,youssef-network-sir-2011}.
Unfortunately, these models are also much more complex.
They utilise a Markov process of individual infection spreads, and in the ultimate master equation description of the process, each configuration has to be treated separately. 
Hence, the total number of equations that we need to solve grows exponentially with respect to the population size.
This problem, known as the \emph{curse of dimensionality} makes such problems prohibitively difficult to solve exactly for networks of moderate and large size.
Therefore, virtually all widely used methods of solving stochastic population models are variants of the Monte Carlo stochastic simulation algorithm~\cite{Gillespie2001}.
Despite its simplicity and popularity, this algorithm is known for its slow convergence following from the central limit theorem.
Alternative approaches include
 mean--field approximations~\cite{keeling-meanfield-1999,rand-meanfield-1999},
 effective degree models~\cite{gleeson-degree-2011,lindquist-degree-2011,kiss-degree-2012},
 and edge--based compartmental models~\cite{miller-edge-2012}, but these models are approximate and rely on truncation of the state space, effects of which on accuracy are difficult to estimate and/or keep below a desired tolerance for a general network.

Recently, a family of \emph{tensor product} methods was proposed for breaking the curse of dimensionality and making high--dimensional problems possible to solve.
In these methods, the solution (in our case, the joint probability distribution function of individual states)
is approximated by a compressed format,
which often converges much faster (e.g. exponentially)
compared to the central limit theorem rate in Monte Carlo methods \cite{griebel2021analysis}.
Starting with basic algorithms for 
  approximating a given array in tensor train (TT)~\cite{osel-tt-2011} 
  or Hierarchical Tucker (HT)~\cite{hk-ht-2009} format,
 new methods were proposed for 
  solving linear systems~\cite{dc-ttgmres-2013,ds-amen-2014}
  and eigenproblems~\cite{dkos-eigb-2014,ds-dmrgamen-2015,rno-gpueig-2019},
  and recently for solving time--dependent problems~\cite{d-tamen-2018}.
Initially motivated by quantum physics~\cite{white-dmrg-1993,klumper-mps-1993,schollwock-2011},
 tensor product algorithms recently extended their domain to a variety of applications, see~\cite{kolda-review-2009,hackbusch-2012,bokh-surv-2015,lars-review-2014}.
In this paper we apply tensor product algorithms to compute, approximately but with controlled accuracy, the joint probability distribution function of network states.

Another difficulty of a general master equation is the exponential number of transitions, in addition to the exponential number of states.
Stochastic population models can often be written as a system of a polynomial number of stochastic chemical reactions, and solved using the chemical master equation (CME)~\cite{vankampen-stochastic-1981}.
In addition to the direct Monte Carlo simulations of the realisations of the model~\cite{Gillespie2001,hemberg-perfect-sampling-2007,AndersonHigham-MLCME-2012,Yates-MLCME-2016},
a direct solution of the CME (or outputs thereof) was proposed using
  an adaptive finite state projection~\cite{munsky-fsp-2006,jahnke-wavelet-cme-2010,Cao-FSP-2016},
  sparse grids~\cite{hegland-cme-2007},
  radial basis functions~\cite{Schuette-RBF-CME-2015},
  neural networks~\cite{Khammash-NN-CME-2021,Grima-CME-NN-2022},
  and tensor product approximations, in particular,
     in the Tucker decomposition~\cite{jahnke-cme-2008},
     CP decomposition~\cite{Ammar-cme-2011,hegland-cme-2011},
     and TT decomposition~\cite{kkns-cme-2014,dkh-cme-2014,d-tamen-2018,Sidje-TT-CME-2017,Dinh-QTT-CME-2020,Ion-TT-CME-2021,Schuette-CME-CO-2016}.
Most of these papers consider the CME formulations of gene regulatory networks, where an accurate description of stochasticity is important due to small copy numbers.
However, there seem to be a little coverage of population models.
Somewhat related is a lattice model of unimolecular adsorption/desorption which was explored in \cite{Schuette-CME-CO-2016}.

In this paper we apply tensor product algorithms to solve the exponentially large systems of ODEs that mathematically capture the evolution of epidemics on networks, without any uncontrollable approximations caused by the simplification of the model.
The fast convergence of the tensor product approximation allows us to solve the CME to extremely high accuracies, up to $6$ decimal digits.
This enables accurate estimation of probabilities of \emph{rare events}, such as simultaneous infection of a large number of people in a network.
Rare event simulation is a infamously formidable task, since the number of samples in a direct Monte Carlo method needs to be inversely proportional to the (small) event probability \cite{botev2008efficient,peherstorfer2018multifidelity,wagner2020multilevel}.
We demonstrate that we can accurately estimate events of probability as small as $10^{-6}$ in a small world network of $50$ individuals.

\section{Background}

The original Kermack--McKendrick model~\cite{kmk-sir-1927} separated people in three groups --- susceptible, infected and recovered --- and  described the state of the epidemics by the size of each group or \emph{compartment}.
For this information to be sufficient for describing the dynamics of epidemic an implied assumption has to be made that the system is \emph{well--mixed} or homogeneous, i.e. each member is in contact with everyone and the disease can spread from each infected person to each susceptible person with the same probability.
This assumption is not very realistic --- the network of contacts between people normally has a complex structure, with some people having (much) more contacts than others.
In this case it is not possible to describe the situation with just specifying sized of all compartments, as the location of infected people in the network plays a key role in the dynamics of epidemic.
The simplest illustration can be 
  a situation when a single infected person is completely isolated from the rest of the network (and infection can not spread), 
  compared to this infected person been connected to all people in the network (and infection can spread rapidly).

Since the compartmental description can not accurately describe epidemic on a general network, we will need to apply stochastic description, by considering for all states the network can reach and describing the evolution of probabilities of these states.
In this section we 
   recall the corresponding mathematical model of SIR epidemics on networks, 
   explain the computational challenges arising due to the exponentially large size of this model, and 
   briefly describe the stochastic simulation algorithm, which avoids the problem and is widely used in practice because of that.

\subsection{Epidemics on networks}
\begin{figure}[t]
   \beginpgfgraphicnamed{ttsir-pic1}  
   \begin{tikzpicture}[x=\linewidth,y=\linewidth]
   \node[lab] at(.25,.43){\textbf{a}};
    \begin{scope}[shift={(0,0)}, x={(45:18mm)},y={(0:25mm)},z={(90:25mm)}, >={Latex[length=1.5ex]}]
     \foreach \x/\a in {0/I,1/R}{
      \foreach \y/\b in {0/S,1/I,2/R}{
       \foreach \z/\c in {0/S,1/I,2/R}{
        \node[state] (\a\b\c) at(\x,\y,\z){};
        \node[above] at(\a\b\c){$\mathrm{\lowercase\expandafter{\a}\lowercase\expandafter{\b}\lowercase\expandafter{\c}}$};
         \begin{scope}[shift={(\a\b\c)},x=1mm,y=1mm]
          \draw[\a] (-2,-.75) circle (1.5);
          \draw[\b] ( 0,-.75) circle (1.5);
          \draw[\c] (+2,-.75) circle (1.5);
         \end{scope}
        }
      }
     }
     \path [postaction={decorate,decoration={text along path, text format delimiters={|}{|},text={|\scriptsize|initial state},text align=center}},shift={(ISS)}] (200:7.5mm) arc (200:340:7.5mm);
     \draw[rec,->] (ISS) to node[scale=.9,left]{$\gamma$} (RSS);
     \draw[inf,->] (ISS) to node[scale=.9,above]{$\beta$}  (IIS);
     \draw[rec,->] (IIS) to node[scale=.9,above]{$\gamma$} (IRS);
     \draw[inf,->] (IIS) to node[scale=.9,left]{$\beta$}  (III);
     \draw[rec,->] (IIS) to node[scale=.9,left]{$\gamma$} (RIS);
     \draw[rec,->] (IRS) to node[scale=.9,left]{$\gamma$} (RRS);
     \draw[rec,->] (RIS) to node[scale=.9,above]{$\gamma$} (RRS);
     \draw[inf,->] (RIS) to node[scale=.9,near start,left]{$\beta$}   (RII);
     \draw[rec,->] (III) to node[scale=.9,left]{$\gamma$} (RII);
     \draw[rec,->] (III) to node[scale=.9,near end,above]{$\gamma$}  (IRI);
     \draw[rec,->] (III) to node[scale=.9,left]{$\gamma$}  (IIR);
     \draw[rec,->] (RII) to node[scale=.9,near start,left]{$\gamma$}  (RIR);
     \draw[rec,->] (RII) to node[scale=.9,near end,above]{$\gamma$} (RRI);
     \draw[rec,->] (IIR) to node[scale=.9,left]{$\gamma$} (RIR);
     \draw[rec,->] (IIR) to node[scale=.9,near end,above]{$\gamma$}  (IRR);
     \draw[rec,->] (RIR) to node[scale=.9,above]{$\gamma$}  (RRR);
     \draw[rec,->] (IRI) to node[scale=.9,left]{$\gamma$}  (RRI);
     \draw[rec,->] (IRI) to node[scale=.9,near end,left]{$\gamma$} (IRR);
     \draw[rec,->] (IRR) to node[scale=.9,left]{$\gamma$} (RRR);
     \draw[rec,->] (RRI) to node[scale=.9,left]{$\gamma$} (RRR);
    \end{scope}

   \node[lab] at(.75,.43){\textbf{b}};
    \begin{scope}[shift={(0.5,0)}, x={(45:18mm)},y={(0:25mm)},z={(90:25mm)}, >={Latex[length=1.5ex]}]
     \foreach \x/\a in {0/I,1/R}{
      \foreach \y/\b in {0/S,1/I,2/R}{
       \foreach \z/\c in {0/S,1/I,2/R}{
        \node[state] (\a\b\c) at(\x,\y,\z){};
        \node[above] at(\a\b\c){$\mathrm{\lowercase\expandafter{\a}\lowercase\expandafter{\b}\lowercase\expandafter{\c}}$};
         \begin{scope}[shift={(\a\b\c)},x=1mm,y=1mm]
          \pgfmathsetmacro\t{90}
          \draw[\a] ({cos(    \t)},{sin(    \t)-1}) circle (1.5);
          \draw[\b] ({cos(120+\t)},{sin(120+\t)-1}) circle (1.5);
          \draw[\c] ({cos(240+\t)},{sin(240+\t)-1}) circle (1.5);
         \end{scope}
        }
      }
     }
     \path [postaction={decorate,decoration={text along path, text format delimiters={|}{|},text={|\scriptsize|initial state},text align=center}},shift={(ISS)}] (200:7.5mm) arc (200:340:7.5mm);
     \draw[rec,->] (ISS) to node[scale=.9,left]{$\gamma$} (RSS);
     \draw[inf,->] (ISS) to node[scale=.9,above]{$\beta$}  (IIS);
     \draw[inf,->] (ISS) to node[scale=.9,left]{$\beta$}  (ISI);
     \draw[rec,->] (IIS) to node[scale=.9,above]{$\gamma$} (IRS);
     \draw[inf,->] (IIS) to node[scale=.9,left]{$2\beta$}  (III);
     \draw[inf,->] (ISI) to node[scale=.9,above]{$2\beta$}  (III);
     \draw[rec,->] (ISI) to node[scale=.9,above]{$\gamma$}  (RSI);
     \draw[rec,->] (ISI) to node[scale=.9,left]{$\gamma$}  (ISR);
     \draw[rec,->] (ISR) to node[scale=.9,above]{$\gamma$}  (RSR);
     \draw[inf,->] (ISR) to node[scale=.9,near end,above]{$\beta$}  (IIR);
     \draw[rec,->] (RSI) to node[scale=.9,near end,left]{$\gamma$}  (RSR);
     \draw[inf,->] (RSI) to node[scale=.9,near end,above]{$\beta$}  (RII);
     \draw[rec,->] (IIS) to node[scale=.9,left]{$\gamma$} (RIS);
     \draw[rec,->] (IRS) to node[scale=.9,left]{$\gamma$} (RRS);
     \draw[inf,->] (IRS) to node[scale=.9,near end,left]{$\beta$} (IRI);
     \draw[rec,->] (RIS) to node[scale=.9,near end,above]{$\gamma$} (RRS);
     \draw[inf,->] (RIS) to node[scale=.9,near start,left]{$\beta$}   (RII);
     \draw[rec,->] (III) to node[scale=.9,left]{$\gamma$} (RII);
     \draw[rec,->] (III) to node[scale=.9,near end,above]{$\gamma$}  (IRI);
     \draw[rec,->] (III) to node[scale=.9,near end,left]{$\gamma$}  (IIR);
     \draw[rec,->] (RII) to node[scale=.9,near start,left]{$\gamma$}  (RIR);
     \draw[rec,->] (RII) to node[scale=.9,near end,above]{$\gamma$} (RRI);
     \draw[rec,->] (IIR) to node[scale=.9,left]{$\gamma$} (RIR);
     \draw[rec,->] (IIR) to node[scale=.9,near end,above]{$\gamma$}  (IRR);
     \draw[rec,->] (RIR) to node[scale=.9,above]{$\gamma$}  (RRR);
     \draw[rec,->] (IRI) to node[scale=.9,left]{$\gamma$}  (RRI);
     \draw[rec,->] (IRI) to node[scale=.9,near end,left]{$\gamma$} (IRR);
     \draw[rec,->] (IRR) to node[scale=.9,left]{$\gamma$} (RRR);
     \draw[rec,->] (RRI) to node[scale=.9,left]{$\gamma$} (RRR);
    \end{scope}
    \begin{scope}[shift={(.25,.5)},x=1em,y=1em]
      \draw[Is] (-5,0)coordinate(p) ellipse (3em and 1em); \node[alice,above,minimum size=2em,opacity=1] at (p){};
      \draw[Ss] (0,0)coordinate(p) ellipse (3em and 1em);  \node[bob,mirrored,above,minimum size=2em,opacity=1] at (p){};
      \draw[Ss] (5,0)coordinate(p) ellipse (3em and 1em);  \node[charlie,above,minimum size=2em,opacity=1] at (p){};
    \end{scope}
    \begin{scope}[shift={(.75,.5)},x=1em,y=1em]
     \draw[Is] (0.0,.75) coordinate(p) ellipse (3em and 1em); \node[charlie,above,minimum size=2em,opacity=1] at (p){};
     \draw[Ss] (-2.5,-.75) coordinate(p) ellipse (3em and 1em); \node[alice,above,minimum size=2em,opacity=1] at (p){};
     \draw[Ss] (2.5,-.75) coordinate(p) ellipse (3em and 1em); \node[bob,mirrored,above,minimum size=2em,opacity=1] at (p){};
    \end{scope}
   \end{tikzpicture}
   \endpgfgraphicnamed
 \caption{Markov chain transitions between network states:
            (a) SIR epidemic on a chain of $N=3$ people;
            (b) SIR epidemic in a fully mixed network of $N=3$ people.
            }
 \label{fig:CME}
\end{figure}
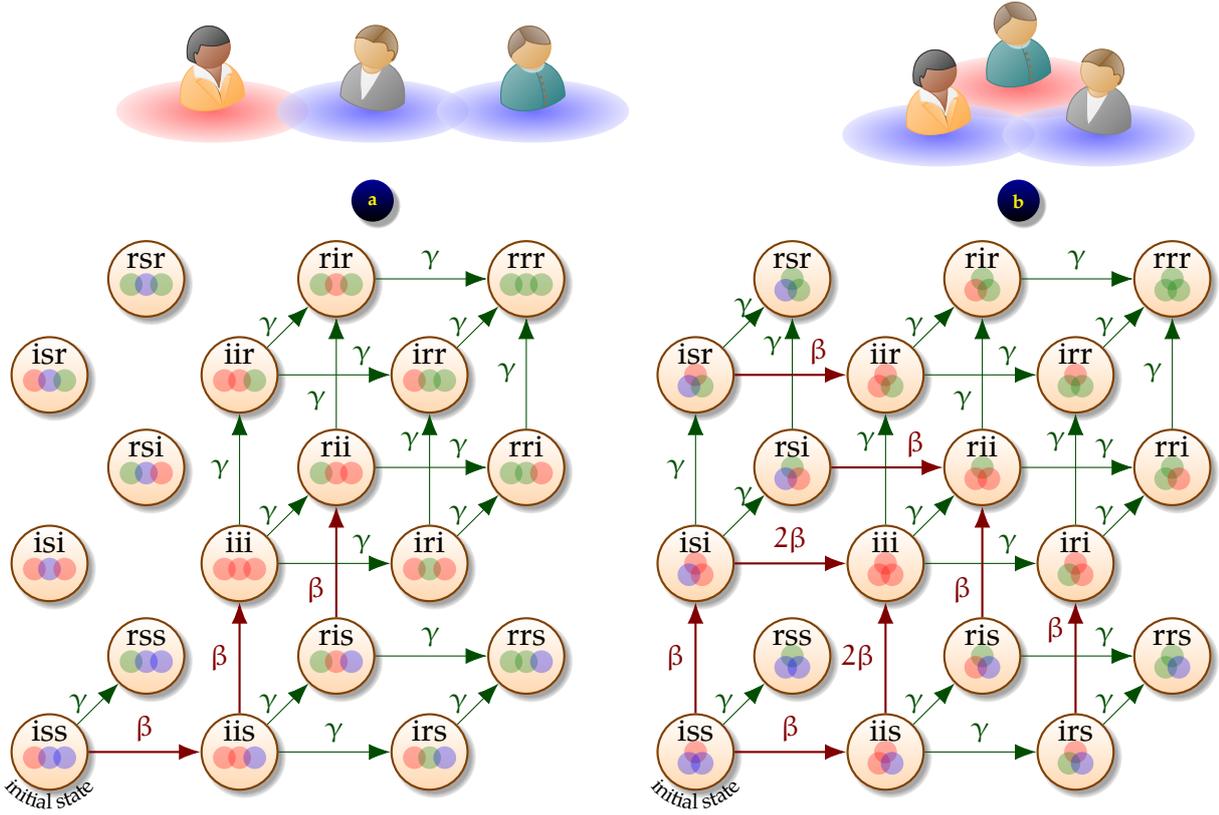

A network (or unweighted simple directed graph) is 
  a set of nodes (vertices, sites) 
  \[
   \Nodes=\{1,\ldots,N\},
  \]
  representing individual people, and 
  a set of links (edges, connections) 
  \[
   \Edges=\{(m,n): \, m\in\Nodes, n\in\Nodes, \, m\neq n\},
  \]
  representing contacts between them.
If $(m,n)\in\Edges,$ which we will also denote using an adjacency relation $m\sim n,$ a person $m$ (if infected) can pass a disease on a person $n$ (if susceptible).
If all contacts can potentially pass disease in both directions (e.g. there are no personal protection measures in place),  the network is undirected, i.e. $(m,n)\in\Edges \iff (n,m)\in\Edges,$ in which case $m\sim n$ is a symmetric relation.

For a SIR epidemic, each person can be in exactly one of three states,
\begin{equation}\label{eq:xn}
 x_n \in \X 
  = \{\mathrm{s}, \mathrm{i}, \mathrm{r}\}
  = \{\text{susceptible}, \text{infected}, \text{recovered}\}
  = \{1,2,3\},
  \qquad \text{for $n\in\Nodes.$}
\end{equation}
Hence, the state of the whole system can be written as
\[
 \x = \begin{pmatrix}x_1 & x_2 & \ldots & x_N \end{pmatrix}^T \in \Omega = \X^N.
\]
We denote the probability to find the system in state $\x$ at time $t$ as 
\( 
 p(\x,t) = p(x_1,x_2,\ldots,x_N,\,t).
\) 
The transitions between the states are described using the two \emph{reactions}, i.e. infection and recovery, respectively: 
\begin{equation}\nonumber
 \begin{split}
  &\Prob\left( \x(t+\delta t) = (x_1,\ldots,x_n=\mathrm{i},\ldots,x_N)  
               \,\middle|\, 
               \x(t) = (x_1,\ldots,x_n=\mathrm{s},\ldots,x_N)
         \right) 
    = I_n(\x) \beta \delta t,
   \\
  &\Prob\left( \x(t+\delta t) = (x_1,\ldots,x_n=\mathrm{r},\ldots,x_N)  
               \,\middle|\, 
               \x(t) = (x_1,\ldots,x_n=\mathrm{i},\ldots,x_N)
         \right) 
    = \gamma \delta t,
 \end{split}
\end{equation}
where 
\(
I_n(\x) = | \{ m\in\Nodes: m\sim n, x_m=\mathrm{i} \} |
\) 
counts the number of infected neighbours of person $n$ in the state $\x.$
These reactions connect the states of the system (as shown in Fig.~\ref{fig:CME}), 
  thus forming a Markov chain network (weighted directed graph with loops), 
  where nodes are network states $\x\in\X^N,$
  links are transitions $\{(\x,\y): \Prob(\text{$\y$ at time $t+\delta t$}\,|\,\text{$\x$ at time $t$})\neq0\},$
  and weights are reaction rates:
\begin{equation}\label{eq:reactions}
 p_{\x\to\y} =
  \begin{cases}
   p_{\x\to\y}^\inf = I_n(\x)\beta, &\text{if $\exists n\in\Nodes:\, x_n=\mathrm{s},\,y_n=\mathrm{i},$ and $x_m=y_m$ for $m\neq n$;} \\  
   p_{\x\to\y}^\rec = \gamma,       &\text{if $\exists n\in\Nodes:\, x_n=\mathrm{i},\,y_n=\mathrm{r},$ and $y_m=x_m$ for $m\neq n$;} \\
    0,           &\text{otherwise.}
  \end{cases}
\end{equation}
Since the probability of the transition depends only on the current state (and not on the history of previous events), this process is a continuous--time Markov chain.
The transition rates describe the dynamics of probabilities of network states as follows:
\begin{equation}\label{eq:cme}
 p'(\x,t) 
  = \sum_{\y\in\Omega} 
        \left( 
          p_{\y\to\x} \cdot p(\y,t) - p_{\x\to\y}\cdot p(\x,t) 
        \right).
\end{equation}
These ODEs need to be solved subject to initial conditions  
\(
p(\x_0,0) = 1
\) 
for the initial state $\x=\x_0$ and 
\(
p(\x,0) = 0
\)
for other states $\x\neq\x_0.$
Collecting all ODEs in a system, we obtain a \emph{Markovian master equation}~\cite{vankampen-stochastic-1981,chen-stoch-sir-2005}, also known as \emph{forward Kolmogorov equations}:
\begin{equation}\label{eq:CME}
 \p'(t) = \A \p(t),\qquad \p'(0)=\p_0,
\end{equation}
where 
\( 
 \p(t) = \begin{bmatrix} p(\x,t) \end{bmatrix}_{\x\in\Omega}
\)
is the unknown \emph{probability distribution function} (p.d.f.),
\(
\p_0
\)
is a unit vector with $1$ in position of the initial state $\x_0,$
and  $\A = [A(\x,\y)]_{\x,\y\in\Omega}$ is matrix with elements
\[
  \begin{cases} 
   A(\x,\x) = -\sum_{\y\in\Omega} p_{\x\to\y}, & \text{on the diagonal} \\
   A(\x,\y) = p_{\y\to\x},                     & \text{off-diagonal, i.e. for $\x\neq\y$}.
  \end{cases}
\]

By solving~\eqref{eq:CME}, we obtain probabilities $p(\x,t)$ for all states $\x\in\Omega,$ and can calculate statistical moments,
\begin{equation}\label{eq:CME.stat}
 \Mean{I(t)} = \sum_{\x\in\Omega} I(\x) p(\x,t), \qquad
  \Var{I(t)} = \sum_{\x\in\Omega} (I(\x) - \Mean{I(t)})^2 p(\x,t),
\end{equation}
where the function 
\(
 I(\x) = I(x_1,\ldots,x_n) = | \{n\in\Nodes: x_n=\mathrm{i} \} |
\)
counts the number of infected individuals for a given state $\x.$

Solving~\eqref{eq:CME} is however not easy due to its large size.
The state space $\Omega=\X^N$ contains $|\Omega|=3^N$ states, meaning that $\p(t)$ is a vector of size $3^N$ and $\A$ is a $3^N \times 3^N$ sparse matrix.
As $N$ increases, the storage and computational costs grow as $\O(3^N)$ and become prohibitively expensive even for modest values $N\gtrsim20.$
This problem, known as the \emph{curse of dimensionality}, is the major obstacle in solving high--dimensional problems that appear in a variety of applications, e.g.
 complex systems, 
 quantum computations, 
 machine learning,
 and epidemics on networks, which we consider in this paper.

\subsection{Simplified models of epidemics on networks}
Since master equations~\eqref{eq:CME} suffer from the curse of dimensionality, a number of alternative approaches to modelling epidemics on networks were developed.

An approach known as \emph{lumping}~\cite{KemenySnell-MC-1976,RogersPitman-MC-1981} attempts to group the network states in classes, in order to obtain a coarse description of the system behaviour by observing transitions between the groups, rather than individual states.
Consider, for example, a network of $N=3$ people fully connected to each other.
A full stochastic description of this network involves $3^N$ network states as shown in Fig.~\ref{fig:CME}(right).
However, we can  collect states $(x_1x_2x_3)$ in groups described by the total number of susceptible and recovered people $(S,R),$ lumping the network model into a stochastic model for SIR epidemics in well--mixed groups.
\begin{flalign*}
 (0,0) & = \{\state{iii}\},
 &
 (1,0) & = \{\state{sii}, \state{isi}, \state{iis}\},
 \\
 (2,0) & = \{\state{ssi}, \state{sis}, \state{iss}\},
 &
 (3,0) & = \{\state{sss}\},
 \\
 (0,1) & = \{\state{rii}, \state{iri}, \state{iir}\}, 
 &
 (1,1) & = \{\state{sir}, \state{sri}, \state{irs}, \state{isr}, \state{rsi}, \state{ris}\},
 \\
 (2,1) & = \{\state{ssr}, \state{srs}, \state{ssr}\},
 &
 (0,2) & = \{\state{irr}, \state{rir}, \state{irr}\},
 \\
 (1,2) & = \{\state{srr}, \state{rsr}, \state{srr}\},
  &
 (0,3) & = \{\state{rrr}\}.
\end{flalign*}
For fully connected network of $N$ people, lumping combines $3^N$ network states in $\tfrac12(N+1)(N+2)=\O(N^2)$ groups, massively reducing the storage and computational complexity, while maintaining all the information necessary for describing the evolution.
This is achieved by sacrificing information about the exact positions of susceptible, infected and recovered people in the network, which can be considered insignificant or unnecessary in this case.
The coarser description obtained by lumping remains exact and perfectly matches the results obtained by the full network description.
However, the effectiveness of lumping depends heavily on the number of symmetries available in the network of connections between people. 
For a fully connected network, permutations of people do not change the structure of connections, hence there are exponentially many symmetries, which explains why lumping is so efficient.
However, even minor modifications of the network such as stepping away from the homogeneous structure of connections, destroy the existing symmetries and make lumping impossible to apply.

Alternative approaches include 
 mean--field approximations~\cite{keeling-meanfield-1999,rand-meanfield-1999}, 
 effective degree models~\cite{gleeson-degree-2011,lindquist-degree-2011,kiss-degree-2012},
 and edge--based compartmental models~\cite{miller-edge-2012}.
These models are approximate and rely on truncation of the state space, effects of which on accuracy are difficult to estimate and/or keep below a desired tolerance for a general network.

In this paper we propose a new method based on tensor product approximation of the CME probability tensor $\p.$
The accuracy of approximations introduced by the proposed method is controlled by a single threshold parameter $\eps$ which can be set accordingly to the desired precision.
The method automatically adjusts parameters controlling the complexity of approximation (so-called \emph{ranks}) to match the desired accuracy.
It does not explicitly rely on network to have symmetries, and thus can be applied to a general network, although the ranks and associated computational costs are network--dependent and may grow uncontrollably for large and densely connected networks.

\subsection{Stochastic simulation algorithm}
In contrast to previous methods, the classical Gillespie's stochastic simulation algorithm (SSA)~\cite{Gillespie77} does not attempt to solve the ODEs~\eqref{eq:CME}.
Instead, it simulates a course of epidemic by sampling random walks $\x(t)$ through the state space $\Omega$ until some desired time $T$.
The sampled trajectories are assumed piecewise--constant, i.e.
\(
x(t) = x_k, 
\)
for $t \in [t_k, t_{k+1}).$
Starting with
\(
k=0, t_0=0, 
\)
and the initial state $x_0,$
SSA uses Monte Carlo sampling to simulate when a next reaction will occur, and which particular reaction will occur.
\begin{enumerate}
 \item Calculate all nonzero transition probabilities (\emph{propensities}) $p_{\x_k\to\y}$ for $\y\in\Omega$.
 \item Calculate the total propensity $p_{\Omega} = \sum_{\y\in\Omega} p_{\x_k\to\y}$.
 \item Sample a time step $\tau$ from exponential distribution with rate parameter $p_{\Omega}$.
 \item Sample a new state $\y_\star \in \Omega$ from the discrete distribution with probabilities $p_{\x_k\to\y} / p_{\Omega}$.
 \item Implement the next step of the walk by setting $\x_{k+1} = \y_\star$, and $t_{k+1}=t_k+\tau$.
 \item If $t_{k+1}<T$, set $k:=k+1$ and repeat from step 1, otherwise stop.
\end{enumerate}
Running this algorithm $\Nssa$ times, we obtain $\Nssa$ sample paths, which can be used to estimate any expectations over the probability distribution defined by the master equation~\eqref{eq:CME}.
Suppose that a quantity of interest $Q(\x,t)$ is a function depending on the state and/or time.
The expectation of $Q$ can be approximated as follows,
\[
 \Mean{ Q(t) }
   \approx \frac{1}{\Nssa} \sum_{s=1}^{\Nssa} Q(\x_k^{(s)}, t_k^{(s)}), 
   \qquad 
   t \in [t_k^{(s)}, t_{k+1}^{(s)}),
\]
where $(t_k^{(s)},\x_k^{(s)})$ represent the $s$--th randomly sampled trajectory.
In other words, the SSA performs a piecewise constant interpolation of the state in time, followed by the Monte Carlo estimator over the sample paths.

Naturally, this estimate contains a statistical error.
Following the central limit theorem, we can conclude that if the variance of the quantity of interest, $\Var{Q}$, is finite, the variance of the estimator of $\Mean{Q}$ is $\Var{Q}/ \Nssa$.
The relative error in the estimate is thus proportional to $(\sqrt{\Var{Q}}/|\Mean{Q}|) \cdot (1/\sqrt{\Nssa})$,
which can be very large if $\sqrt{\Var{Q}} \gg | \Mean{Q} |.$
To compensate for this, a very large $\Nssa$ is needed, which leads to enormous computational costs.
This happens for example in estimation of probabilities of rare events.
In this case $Q(\x,t)$ is an indicator function of the event of interest, with $\Mean{Q} \ll 1$.

Alternative algorithms include for example Tau-Leaping \cite{Gillespie2001} and multi-level simulations \cite{AndersonHigham-MLCME-2012,Yates-MLCME-2016}.
The Tau-Leaping method fixes a time step $\tau$,
and samples (possibly several) reactions within this time step from a Poisson distribution.
Clearly, the pre-selected time step $\tau$ can be larger than the time steps sampled by SSA, which requires fewer steps in total.
However, Tau-Leaping samples biased trajectories, with the bias increasing with $\tau$ \cite{Anderson-tau-leaping-2011}.
Multi-level algorithms allow one to compensate for more time steps resulting from a small $\tau$ by sampling less trajectories, and vice versa.
This alleviates the problem of sampling fast reactions with small time steps.
However, these methods may still struggle with high variance of the quantity of interest, $\sqrt{\Var{Q}} \gg |\Mean{Q}|$.

\section{Methods}

In this section we introduce tensor product approach to solving CME for epidemics on network.

\subsection{Chemical master equation for the network SIR model}
The matrix~$\A$ in the master equation~\eqref{eq:CME} has exponentially large size $3^N\times 3^N,$ which makes classical algorithms struggle from the curse of dimensionality.
Fortunately, it has a hidden tensor product structure, which we will reveal and exploit to solve the problem using tensor product algorithms.

Firstly, note that the right--hand side in~\eqref{eq:cme} contains sums over $|\Omega|=3^N$ states $\y.$
However, for a given state $\x\in\Omega$ most of the transitions $\x\to\y$ and $\y\to\x$ are impossible, i.e. $p_{\x\to\y}=0$ and $p_{\y\to\x}=0.$
We will rewrite sums in a more explicit form by keeping only possible transitions.

From now on we will denote the states of individual nodes~\eqref{eq:xn} using numbers,
\(
x_n \in \{1,2,3\}.
\)
Note that an infection $\x\to\y$ which makes a susceptible person $x_n=1$ infected $y_n=2$ can be written as $\y=\x+\e_n$ where $\e_n\in\Real^N$ is the $n$-th unit vector.
A recovery $\x\to\y$ which makes an infected person $x_n=2$ recovered $y_n=3$ also can be written as $\y=\x+\e_n.$
Hence, both the infection and recovery reactions are described by the \emph{stoichiometry} $1$.
This means that  
  $p_{\y\to\x}=0$ unless $\exists n\in\Nodes:\, \x=\y+\e_n,$ and
  $p_{\x\to\y}=0$ unless $\exists n\in\Nodes:\, \y=\x+\e_n.$
Hence we can rewrite~\eqref{eq:cme} as follows
\begin{equation}\label{eq:cme.1}
 \begin{split}
   p'(\x,t) 
     & = \sum_{n=1}^N \left[\strut p_{\y\to\x} \cdot p(\y,t) \right]_{\x=\y+\e_n}
       - \sum_{n=1}^N \left[\strut p_{\x\to\y} \cdot p(\x,t) \right]_{\y=\x+\e_n}
   \\& = \sum_{n=1}^N \underbrace{p_{(\x-\e_n)\to\x}}_{a_n(\x-\e_n)} \cdot p(\x-\e_n,t)
       - \sum_{n=1}^N \underbrace{p_{\x\to(\x+\e_n)}}_{a_n(\x)} \cdot p(\x,t),
 \end{split}   
\end{equation}
keeping only $N$ terms in each sum and introducing notation $a_n(\x)=p_{\x\to(\x+\e_n)}$ for reaction rates of stoichiometry $1.$
This form of the master equation is often called \emph{chemical master equation} (CME) following~\cite{Gillespie2001}.
As shown in~\eqref{eq:reactions}, the specific formula for the reaction rate $a_n(\x)$ depends on whether this reaction is infection or recovery, which in turn depends on the value of $x_n.$
Using an \emph{indicator} function 
\[
 \Indicator{\text{\emph{condition}}}
  = \begin{cases}
   1, & \text{if \emph{condition} is true} \\
   0, & \text{if \emph{condition} is false},
  \end{cases}
\]
we rewrite~\eqref{eq:reactions} as follows:
\begin{equation}\label{eq:a}
 \begin{split}
  a_n(\x) = p_{\x\to(\x+\e_n)} 
      & = \Indicator{x_n=1} \cdot p^\inf_{\x\to(\x+\e_n)} 
        + \Indicator{x_n=2} \cdot p^\rec_{\x\to(\x+\e_n)} 
  \\  & = \Indicator{x_n=1} \cdot I_n(x) \beta
        + \Indicator{x_n=2} \cdot \gamma
  \\  & = \sum_{m\sim n}\underbrace{\beta \Indicator{x_n=1} \Indicator{x_m=2} }_{a_{m\to n}^\inf(\x)}
        + \underbrace{\gamma \Indicator{x_n=2}}_{a_n^\rec(\x)},
 \end{split}
\end{equation}
where 
\(
I_n(\x) = \sum_{m\sim n} \Indicator{x_m=2}
\) 
counts infected neighbours of person $n.$

All ODEs~\eqref{eq:cme.1} taken together form the master equation with the vector of unknowns 
\(
\p(t)=[p(\x,t)]_{\x\in\Omega}.
\)
We will place the probability of the state $x=(x_1,x_2,\ldots,x_N)$ in position 
\(
\overline{x_1x_2\ldots x_N} 
 = 3^{N-1}(x_1-1) 
 + 3^{N-2} (x_2-1) 
 + \cdots
 + 3^0 x_N
\)
of vector $\p(t).$
With this \emph{big-endian} ordering, vector $\a_n^\rec=[a_n^\rec(\x)]_{\x\in\Omega}$ from~\eqref{eq:a} can be written as
\begin{equation}\label{eq:arec}
 \a_n^\rec 
  = \gamma 
    \evec \otimes \cdots \otimes \evec 
    \otimes \ivec \otimes 
    \evec  \otimes \cdots \otimes \evec,
\end{equation}
where 
   $\ivec=\begin{pmatrix}0&1&0\end{pmatrix}^T$ appears in position $n,$ 
   $\evec=\begin{pmatrix}1&1&1\end{pmatrix}^T$ appear in all positions $1,\ldots,N$ except $n,$
    and $\otimes$ denotes Kronecker (tensor) product.\footnote{Recall that the Kronecker product $C = A \otimes B$ for $A \in \Real^{p \times q}$ and $B \in \Real^{m\times n}$ is a $pm \times qn$ matrix with elements $C(i+(j-1)m, k+(\ell-1)n) = A(j,\ell) B(i,k)$. The Kronecker product is distributive and associative.}
Similarly, vector $\a_{m\to n}^\inf=[a_{m\to n}^\inf(\x)]_{\x\in\Omega}$ from~\eqref{eq:a} can be written as
\begin{equation}\label{eq:ainf}
 \a_{m\to n}^\inf 
  = \beta 
   \evec \otimes \cdots \otimes \evec 
    \otimes \svec \otimes 
    \evec  \otimes \cdots \otimes \evec
    \otimes \ivec \otimes 
    \evec  \otimes \cdots \otimes \evec,
\end{equation}
where 
   $\svec=\begin{pmatrix}1&0&0\end{pmatrix}^T$ appears in position $n,$ 
   $\ivec=\begin{pmatrix}0&1&0\end{pmatrix}^T$ appear in positions $m\sim n,$ 
   $\evec=\begin{pmatrix}1&1&1\end{pmatrix}^T$ appear elsewhere.
Note that 
  appearance of basis vectors $\svec$ and $\ivec$ realises conditions of indicator functions $\Indicator{x_n=1}$ and $\Indicator{x_m=2}$ in~\eqref{eq:a},
  and $\evec$ appears in positions of nodes not affected by any conditions. 
Now the vector $[a_n(\x) p(\x,t)]_{\x\in\Omega}$ in the second term of~\eqref{eq:cme.1} can be written as
\begin{equation}\label{eq:a2}
  \begin{split}
   \diag(\a_n) \p
     & = \left( \sum_{m \sim n} \diag(\a^\inf_{m\to n}) + \diag(\a^\rec_n) \right) \p,
   \\
   \diag(\a^\inf_{m\to n})  & = 
                      \beta\cdot
                       \Id \otimes \cdots \otimes \Id 
                       \otimes \diag(\svec) \otimes 
                       \Id  \otimes \cdots \otimes \Id
                       \otimes \diag(\ivec) \otimes 
                       \Id  \otimes \cdots \otimes \Id,
   \\
   \diag(\a^\rec_n)  & = 
                       \gamma\cdot
                       \Id \otimes \cdots \otimes \Id 
                       \otimes \diag(\ivec) \otimes 
                       \Id  \otimes \cdots \otimes \Id,
  \end{split}
\end{equation}
where $\Id = \diag(\evec)$ is a $3\times 3$ identity matrix.

The first term in the right--hand side of~\eqref{eq:cme.1} contains the probability of the shifted state 
\(
p(\x-\e_n,t)
\)
which we can express in terms of probabilities $p(\y,t)$ as follows
\begin{equation}\label{eq:shift}
 \begin{split}
  q(\x,t) 
    & = p(\x-\e_n,t) 
  \\& = \sum_{\y\in\Omega}
             \Indicator{x_1=y_1} \cdots
             \Indicator{x_{n-1}=y_{n-1}} \cdot 
             \Indicator{x_n-1=y_n} \cdot 
             \Indicator{x_{n+1}=y_{n+1}} \cdots
             \Indicator{x_N=y_N} 
             \cdot p(\y,t),
             \\
   \q(t) & =  
             \underbrace{\left(
             \Id \otimes \cdots \otimes \Id \otimes J^T \otimes \Id \otimes \cdots \otimes \Id
             \right)}_{\J_n^T}
             \p(t),
 \end{split}
\end{equation}
where the \emph{shift matrix}
  \(
  J^T=\left(\begin{smallmatrix} \zero& \zero& \zero \\ 1 &\zero & \zero\\ \zero& 1 &\zero \end{smallmatrix}\right)
  \)
  appears in position $n,$ 
  and the identity matrix
  \(
  \Id=\left(\begin{smallmatrix} 1 & \zero & \zero \\ \zero & 1 & \zero \\ \zero & \zero & 1\end{smallmatrix}\right)
  \) 
  appears elsewhere.
Because of this special structure, we can say that the $3^N \times 3^N$ matrix $\J^T_n$ acts on $n$--th site of the system only.

The same process can be applied to the vector $[a_n(\x-\e_n)p(\x-\e_n,t)]_{\x\in\Omega}$ in the first term of~\eqref{eq:cme.1}, which gives
\begin{equation}\nonumber
  \p' = \sum_{n=1}^N \J_n^T \diag(\a_n) \p - \sum_{n=1}^N \diag(\a_n) \p.
\end{equation}
Plugging in the tensor product expansion~\eqref{eq:a2}, we obtain the matrix of the master equation \eqref{eq:CME} as
\begin{equation} \label{eq:CME.TT}
 \begin{split}
 \A  
   & = \sum_{n=1}^N \sum_{m\sim n} \beta \cdot \Id \otimes \cdots \otimes \Id \otimes (J^T-\Id) \diag(\svec) \otimes \Id \otimes \cdots \otimes \Id \otimes \diag(\ivec) \otimes \Id \otimes \cdots \otimes \Id
   \\ & + \sum_{n=1}^N \gamma \cdot \Id \otimes \cdots \otimes \Id \otimes (J^T-\Id) \diag(\ivec) \otimes \Id \otimes \cdots \otimes \Id.
 \end{split}
\end{equation}

The special \emph{tensor product} form of matrices $\J_n^T$ in~\eqref{eq:shift} and $\A$ in~\eqref{eq:CME.TT} allows us to define large $3^N \times 3^N$ matrices as a tensor product of small $3 \times 3$ matrices acting on individual sites of the system.
By \emph{defining} matrices is this form we can avoid \emph{computing} them explicitly, hence reducing storage requirements significantly.
For example, a $3^N \times 3^N$ matrix $\J_n^T$ has $2 \cdot 3^{N-1}$ nonzero elements, hence storing it in full (but sparse) form requires $\O(3^N)$ memory. 
However the factors of tensor product in~\eqref{eq:shift} have $3N-1$ nonzero elements in total, hence we can keep the factorised matrix $\J_n^T$ using $\O(3N)$ storage.
Similarly, matrix $\A$ in \eqref{eq:CME.TT} is represented by $(|\Edges|+N)$ tensor product terms, reducing total storage from $\O(3^N)$ to $\O(3|\Edges|N+3N^2)=\O(3 (\langle k \rangle +1)N^2),$ where $\langle k \rangle = |\Edges|/|\Nodes|$ is the average degree of the network.

\subsection{Tensor product factorisations}
Maintaining the factorised tensor product form for the matrix $\A$ of the chemical master equation~\eqref{eq:CME.TT}, we remove the \emph{curse of dimensionality} for storage of $\A.$
To similarly reduce the storage and computational costs for the unknown probability distribution function $\p(t)$ and make the numerical solution possible, we need to assume a similar tensor product representation for $\p(t)$ to hold exactly or approximately with sufficiently good accuracy.
For the sake of simplicity, let's first drop the dependency on $t$ and consider a $N$-tensor $\p=[p(x_1,x_2,\ldots,x_N)]$ of size $3 \times 3 \times \cdots \times 3.$
The simplest attempt would be to factorise $\p$ mimicking the sum of tensor products in \eqref{eq:CME.TT} with some $R$ terms,
\begin{equation}\label{eq:cp}
 \p \approx \tilde\p
     = \sum_{\alpha=1}^R \p^{[1]}_{\alpha} \otimes \cdots \otimes \p^{[N]}_{\alpha},
\end{equation}
with some $\p^{[k]}_{\alpha} \in \Real^3.$
This decomposition is called \emph{canonical polyadic} (CP) format \cite{hitchcock-sum-1927,kolda-review-2009}.
If all high--dimensional tensors are kept in tensor product format, all computations can be performed with one--site factors instead of full vectors and matrices, lifting the curse of dimensionality.
Unfortunately, the CP format~\eqref{eq:cp} can be unstable and the best approximation does not always exist~\cite{desilva-2008}, which makes it less attractive.

A more structured representation that admits stable computations is the \emph{tensor train} (TT) decomposition~\cite{osel-tt-2011}.
A vector $\p\in\Real^{3^N}$ is said to be approximated in a TT decomposition with a relative error threshold $\eps \geq 0$ if there exist \emph{TT cores} 
\(
\p^{(n)}_{\alpha_{n-1},\alpha_n} \in \Real^3,
\)
$n=1,\ldots,N,$ such that
 \begin{equation}\label{eq:tt}
  \p \approx \tilde\p 
      = \sum_{\alpha_0,\ldots,\alpha_N=1}^{r_0,\ldots,r_N} 
             \p^{(1)}_{\alpha_0,\alpha_1} 
             \otimes \cdots \otimes 
             \p^{(n)}_{\alpha_{n-1},\alpha_n} 
             \otimes \cdots \otimes 
             \p^{(N)}_{\alpha_{N-1},\alpha_N},
  \qquad
 \end{equation}
with $\|\p - \tilde\p\|_2 \leq \eps \|\p\|_2.$
The ranges of the summation indices $r_0,\ldots,r_N$ are called \emph{TT ranks}.
Each core $\p^{(n)}$ contains information related to person $n$ in the network, and the summation indices $\alpha_{n-1},\alpha_n$ of core $\p^{(n)}$ link it to cores $\p^{(n-1)}$ and $\p^{(n+1)}.$
This linear arrangement of tensor train cores explain the name of the format.

In contrast to the CP format, the TT ranks $r_n$ are the ranks of \emph{unfolding} matrices,
\(
r_n = \rank\begin{bmatrix}\tilde p(\overline{x_1\ldots x_n}, \overline{x_{n+1}\ldots x_N})\end{bmatrix}.
\)
This ensures the existence of best approximation in the TT format with the TT ranks
\(
r_n \leq \min(3^n, 3^{N-n}),
\)
for $n=0,\ldots,N$.
In particular, we have $r_0=r_N=1$.
If $\p$ were a probability distribution function for independent random variables, we would have
\(
p(x_1,\ldots,x_d) = p^{(1)}(x_1) \cdots p^{(d)}(x_d),
\)
hence 
\( 
r_0=r_1=\cdots = r_N=1.
\)
On the other side, a \emph{strongly correlated} distribution may have large TT ranks, potentially approaching their (exponentially large) upper bounds, in which case the TT approximation won't be effective.
Our approach is aimed at \emph{weakly correlated} distributions, such as those with all TT ranks bounded,
\(
r=\max_{n=1,\ldots,N-1} r_n \ll 3^N.
\)
For example, existence of TT approximations with $r = \O(N)$ was proven for stationary distributions describing mass-action and Michaelis--Menten kinetics~\cite{kkns-cme-theory-2015}.
In this case, it is sufficient to store only $\O(N r^2)$ elements of the TT cores to encode the entire vector $\tilde\p$.
Moreover, the pairwise structure of the summation over $\alpha_n$, resembling the dyadic factorisation of $\tilde \p$ written as a matrix,
offers a stable way to compute the TT approximation for any vector by using the (truncated) singular value decomposition (SVD) as shown in~\cite{osel-tt-2011}.
Alternatively, the TT approximations can be computed by alternating least square optimisation over the TT cores as shown in~\cite{holtz-ALS-DMRG-2012} and also in earlier work on matrix product states (MPS)~\cite{fannes-mps-1992,klumper-mps-1993} in context of quantum physics.

\subsection{Discretisation in time}
We discretise the dynamical system~\eqref{eq:CME} on 
\(
t \in [0,T],
\)
where $T$ is the desired time horizon.
We introduce a set of \emph{reference} time points 
\(
\{t_k\}_{k=1}^K,
\)
such that
\(
0=t_0 < t_1 < \cdots < t_K=T.
\)
These can be the points where the solution is ultimately sought, or they can be determined adaptively to control the discretisation error~\cite{byrne-ode-1975}.
On each subinterval 
\(
(t_{k-1}, t_k]
\)
we introduce a basis of Lagrange polynomials 
\(
\{\phi_\ell^{(k)}(t)\}_{\ell=1}^L
\)
centred at Chebysh\"ev nodes 
\begin{equation}\label{eq:cheb}
 t_\ell^{(k)} = \tfrac12 (t_{k-1}+t_k) + \tfrac12(t_k-t_{k-1}) \cos\left(\pi\, \tfrac{\ell-1}{L}\right),
  \qquad 
 \ell = 1,\ldots,L.
\end{equation}
The p.d.f. can now be approximated as
\[
\p(t) 
  \approx \sum_{\ell=1}^L  \p(t_\ell^{(k)}) \cdot \phi_\ell^{(k)}(t), 
\qquad 
t \in (t_{k-1}, t_k],
\]
and the spectral approximation theory guarantees an exponential convergence in $L$~\cite{trefethen-spectral-2000} if $\p(t)$ is analytic on $(t_{k-1}, t_k].$
On each subinterval we want to compute all values $p(\x, t_\ell^{(k)}),$ forming a vector
\(
\bar\p^{(k)} = [p(x_1,x_2,\ldots,x_N,\,t_\ell^{(k)})] \in \Real^{3^N L},
\)
which can be also seen as a tensor with $(N+1)$ modes and dimensions $3 \times 3 \times \cdots \times 3 \times L.$
The TT decomposition \eqref{eq:tt} is expanded accordingly:
\begin{equation}\label{eq:ttt}
 \bar\p^{(k)} 
    \approx  \tilde\p^{(k)}
    = \sum_{\alpha_0,\ldots,\alpha_{N+1}=1}^{r_0,\ldots,r_{N+1}} 
         \p^{(k,1)}_{\alpha_0,\alpha_1} 
         \otimes \cdots \otimes 
         \p^{(k,n)}_{\alpha_{n-1},\alpha_n} 
         \otimes \cdots \otimes 
         \p^{(k,N)}_{\alpha_{N-1},\alpha_N} \otimes 
         \p^{(k,N+1)}_{\alpha_N,\alpha_{N+1}},
\end{equation}
where now 
  $r_{N+1}=1$, 
  $r_N\geq 1$ (in general), and 
  the new TT core $\p^{(k,N+1)}_{\alpha_N,\alpha_{N+1}} \in \Real^L$ encodes the dependence on time.
This p.d.f. can now be interpolated at any $t\in(t_{k-1},t_k]$ as
\begin{equation}\label{eq:ttt-cont}
  \p(t) 
    \approx \tilde\p(t)
    = \sum_{\alpha_0,\ldots,\alpha_{N+1}} 
           \p^{(k,1)}_{\alpha_0,\alpha_1} 
           \otimes \cdots \otimes 
           \p^{(k,n)}_{\alpha_{n-1},\alpha_n} 
           \otimes \cdots \otimes 
           \p^{(k,N)}_{\alpha_{N-1},\alpha_N} 
           \cdot 
           \left(
                \sum_{\ell=1}^L
                     \p^{(k,N+1)}_{\alpha_N,\alpha_{N+1}}(\ell) 
                     \cdot 
                     \phi_\ell^{(k)}(t)
                \right).
\end{equation}

The time derivative is replaced by a \emph{differentiation matrix}
\(
D^{(k)} = [(\phi_{\ell'}^{(k)})'(t_\ell^{(k)})]_{\ell,\ell'=1}^L.
\)
This allows us to propagate the master equation~\eqref{eq:CME} through the interval $(t_{k-1},t_k]$ by solving a linear equation
\begin{equation} \label{eq:timescheme}
 \underbrace{\left(\Id_{3^N} \otimes D^{(k)} - \A \otimes \Id_L \right)}_{\A^{(k)}} \tilde\p^{(k)}
  =
  \underbrace{\p(t_{k-1}) \otimes (D^{(k)} \e_L)}_{\f^{(k)}},
\end{equation}
where 
  $\Id_n$ is $n\times n$ identity matrix, and
  $\e_n$ is the vector of all ones of size $n.$
The initial state $\p(t_{k-1})$ 
  is taken as the initial condition $\p(0)$ if $k=1$ 
  or interpolated from the previous subinterval using \eqref{eq:ttt-cont}.
Plugging in \eqref{eq:CME.TT} and noticing that 
\(
\Id_{3^N} = \Id \otimes \cdots \otimes \Id,
\)
the matrix $\A^{(k)}$ can be also written in a tensor product form similar to \eqref{eq:CME.TT} but with one extra term.
In the same way, if $\p(t_{k-1})$ is replaced by the TT decomposition \eqref{eq:ttt-cont},
the right hand side $\f^{(k)}$ can be written as a TT decomposition as well.

\subsection{Tensor product algorithms for solving linear systems}
To solve the linear system
\eqref{eq:timescheme} we can use the Alternating Linear Scheme (ALS) algorithm \cite{holtz-ALS-DMRG-2012}, or the more robust
Alternating Minimal Energy (AMEn) algorithm \cite{ds-amen-2014}.
The basic ALS algorithm solves~\eqref{eq:timescheme} 
  by iterating over $n=1,\ldots,N+1$, 
  fixing in each step all TT cores in \eqref{eq:ttt} but $\p^{(k,n)}$, 
  and solving the resulting over-determined system for the elements of $\p^{(k,n)}$.
This can be seen by stretching the TT core
\(
\p^{(k,n)}
\)
into a long vector 
\(
\mathrm{p}^{(k,n)} 
  = \mathop{\mathrm{vec}}(\p^{(k,n)})
  = [\p^{(k,n)}_{\alpha_{n-1},\alpha_n}(x_n)]
\)
and introducing a \emph{frame} matrix
\[
\Proj_{\neq n}^{(k)}
 = \left(\sum_{\alpha_0\ldots\alpha_{n-2}} 
      \p^{(k,1)}_{\alpha_0,\alpha_1}
      \otimes \cdots \otimes 
      \p^{(k,n-1)}_{\alpha_{n-2},:}
   \right)
        \otimes \Id \otimes
   \left(\sum_{\alpha_{n+1}\ldots\alpha_{N+1}} 
       \p^{(k,n+1)}_{:,\alpha_{n+1}}
       \otimes \cdots \otimes 
       \p^{(k,N+1)}_{\alpha_{N},\alpha_{N+1}}
    \right),
\]
which is of size $3^N L \times 3 r_{n-1} r_n$ for $n\leq N,$ and of size $3^N L \times r_N L$ for $n=N+1.$
Now the TT decomposition \eqref{eq:ttt} can be written as a linear map
\(
\tilde \p^{(k)} = \Proj_{\neq n}^{(k)} \mathrm{p}^{(k,n)}.
\)
The ALS method performs the Galerkin projection to solve a reduced linear system
\begin{equation}\label{eq:localsys}
 ((\Proj_{\neq n}^{(k)})^T \A^{(k)} \Proj_{\neq n}^{(k)})\mathrm{p}^{(k,n)} 
    = (\Proj_{\neq n}^{(k)})^T \f^{(k)}
\end{equation}
subsequently for $n=1,\ldots,N+1.$
This system can be assembled and solved efficiently~\cite{holtz-ALS-DMRG-2012} due 
  $\Proj_{\neq n}^{(k)}$, $\A^{(k)}$ and $\f^{(k)}$ available in TT format.
The total complexity is $\O(Nr^3).$

However, this simple ALS method has two drawbacks:
  TT ranks are fixed from the beginning (and may not match the ranks required for the unknown solution), and
  the sequential optimisation process may stuck in a too inaccurate solution.

The AMEn method \cite{ds-amen-2014} circumvents these issues by computing also a TT approximation of the residual
\(
 \tilde\z^{(k)} \approx\z^{(k)} = \f^{(k)} - \A^{(k)}\tilde\p^{(k)}.
\)
This is done effectively by minimising the error 
\(
\|\tilde\z^{(k)} - \z^{(k)}\|_2^2,
\)
using standard alternating least squares algorithm~\cite{kolda-review-2009,holtz-ALS-DMRG-2012},
and by expanding the search space 
\(
\Proj_{\neq n}^{(k)}
\)
with a TT core $\z^{(k,n-1)}$ of the residual $\tilde\z^{(k)},$
\[
\Proj_{\neq n}^{(k)}
 = \left(\sum_{\alpha_0\ldots\alpha_{n-2}} 
     \p^{(k,1)}_{\alpha_0,\alpha_1}
     \otimes \cdots \otimes 
     \begin{bmatrix}
        \p^{(k,n-1)}_{\alpha_{n-2},:} 
        & 
        \z^{(k,n-1)}_{\alpha_{n-2},:}
      \end{bmatrix}
    \right)
        \otimes \Id \otimes
   \left(\sum_{\alpha_{n+1}\ldots\alpha_{N+1}} 
         \p^{(k,n+1)}_{:,\alpha_{n+1}}
         \otimes \cdots \otimes 
         \p^{(k,N+1)}_{\alpha_N,\alpha_{N+1}}
      \right).
\]
This allows one to expand the TT rank $r_{n-1}$ if it was underestimated.
On the other hand, truncating the singular values of $\p^{(k,n-1)}$ below the desired error threshold~\cite{osel-tt-2011}, one can reduce TT ranks if they are overestimated.
Moreover, inception of the global residual information 
 ensures global convergence of the AMEn algorithm~\cite{ds-amen-2014} to the solution of~\eqref{eq:timescheme} 
 and enhances its convergence rate in practical computations.

\subsection{Time step adaptation for local error control}
Finally, the pseudospectral time discretisation \eqref{eq:timescheme}, combined with the TT decomposition \eqref{eq:ttt}, allows one to estimate the time discretisation error in a computationally efficient way,
and hence to adapt the reference time points $t_k$ using standard local error control methods.
We check the accuracy of the computed solution $\tilde\p^{(k)},$ on a finer Chebysh\"ev grid
\(
\hat t_\ell^{(k)} \in (t_{k-1},t_k],
\)
given by~\eqref{eq:cheb} with $2L$ nodes.


We 
  construct the linear system~\eqref{eq:timescheme} on the finer grid $\{\hat t_\ell^{(k)}\}_{\ell=1}^{2L},$ 
  plug in the solution $\tilde\p(t),$ interpolated from the current grid to the finer grid using~\eqref{eq:ttt-cont}, 
  and evaluate the residual, which can be done efficiently due to the TT and Kronecker product structures.
If the residual norm exceeds the desired error threshold, 
  the current solution is rejected, 
  the time step is reduced, and 
  the solution is recomputed on a smaller interval $(t_{k-1}, t_k].$
If the residual norm is well below the desired efficiency threshold, the size of the next time interval is increased.
This error control mechanism is implemented in the tAMEn (time--dependent AMEn) algorithm~\cite{d-tamen-2018},
which we use in the numerical experiments.

\subsection{Evaluation of observables}

When the p.d.f. $\p(t)$ is computed in the TT format,
we can evaluate observables, such as $\Mean{I(t)}$ and $\Var{I(t)}$ in~\eqref{eq:CME.stat}.
However, we need to avoid taking sums over $3^N$ network states $\x\in\Omega$ and use more efficient strategy exploiting the properties of the TT format~\eqref{eq:tt}.

As an introductory example, suppose that we solved CME~\eqref{eq:CME} and obtained $\p(t)$ in TT format~\eqref{eq:ttt}.
The representation $\p(t)$ is essentially a sequence of arrays~\eqref{eq:ttt} spanning intervals $(t_{k-1},t_k].$
It is therefore sufficient to discuss how to calculate observables at a particular time $t,$ and then repeat the procedure for all intervals, thus covering the entire integration region $(0,T].$
Suppose we obtained $\p(t)\approx\tilde\p$ in the TT format~\eqref{eq:tt}.
We may be interested in calculating total probability $\sum_{\x\in\Omega}\tilde p(\x),$ which equals $1$ in theory, but may slightly deviate due to discretisation errors during numerical integration of~\eqref{eq:CME} and approximation of the solution in TT format.
\begin{equation}\label{eq:Ptot}
 \begin{split}
  |\tilde\p|_1 
      & = \sum_{\x\in\Omega} \tilde p(\x) 
        = \sum_{\x\in\X^N} \tilde p(x_1,x_2,\ldots,x_N)
     \\ & = \sum_{\alpha_1} \left[\left(\sum_{x_1\in\X}\p^{(1)}_{\alpha_{1}}(x_1)\right) \cdots
     \sum_{\alpha_{N-1}} \left(\sum_{x_{N-1}\in\X}\p^{(N-1)}_{\alpha_{N-2},\alpha_{N-1}}(x_{N-1})\right)
     \left(\sum_{x_N\in\X}\p^{(N)}_{\alpha_{N-1}}(x_N)\right) \right]
 \end{split}
\end{equation}
Taking unimodal sums inside round brackets costs $\O(N r^2)$ operations, after which we need to sum over $\alpha_{N-1},\ldots,\alpha_1$, which can be implemented as a sequence of matrix products~\cite{osel-tt-2011}.
This takes another $\O(N r^2)$ operations, where $r$ denotes the largest TT rank of $\tilde\p.$
Hence, the total complexity is no longer exponential, but polynomial in number of people in the network.
Rescaling the p.d.f. $\p = \tilde\p / |\tilde\p|_1$ we can (partly) mitigate the errors introduced during computations.

Now let's consider computing the mean
\[
 \Mean{ I }
 = \sum_{\x\in\Omega} I(\x) p(\x)
 = | [ I(\x) p(\x) ]_{\x\in\Omega} |_1
 = | [ I(\x) ]_{\x\in\Omega} \odot  \p  |_1
 = \langle [I(\x)]_{\x\in\Omega}, \p \rangle,
\]
where $\odot$ denotes the Hadamard (pointwise) product of vectors, matrices or tensors,
  $|\,\cdot\,|_1$ denotes sum over all elements,
  and $\langle \cdot,\cdot \rangle$ denotes scalar product.
Similarly to~\eqref{eq:arec} and~\eqref{eq:ainf}, we obtain
\begin{equation}\label{eq:I.cp}
 \begin{split}
   I(\x) 
    & = I(x_1,x_2,\ldots,x_N) 
      = \Indicator{x_1 = \mathrm{i}}
      + \Indicator{x_2 = \mathrm{i}}
      + \cdots 
      + \Indicator{x_N = \mathrm{i}}, \\
   [I(\x)]_{\x\in\Omega} 
    & = \ivec \otimes  \evec \otimes \cdots \otimes \evec
      + \evec \otimes \ivec  \otimes \cdots \otimes \evec
      + \cdots
      + \evec \otimes \evec  \otimes \cdots \otimes \ivec.
 \end{split}
\end{equation}
We note that the tensor $[I(\x)]_{\x\in\Omega}$ admits CP decomposition~\eqref{eq:cp} with tensor rank $R=N.$
However, due to a special structure of the rank-one terms the corresponding TT decomposition has TT ranks all equal to two:
\begin{equation}\label{eq:I.tt}
 \begin{split}
  [I(\x)]_{\x\in\Omega} 
    & = \sum_{\alpha_1,\ldots,\alpha_{N-1}=1}^{2} 
       \begin{bmatrix} \evec & \ivec \end{bmatrix}_{\alpha_1} 
        \otimes
        \begin{bmatrix} \evec & \ivec \\ \zero & \evec \end{bmatrix}_{\alpha_1,\alpha_2} 
        \otimes\cdots\otimes 
        \begin{bmatrix} \evec & \ivec \\ \zero & \evec \end{bmatrix}_{\alpha_{N-2},\alpha_{N-1}} 
         \otimes
        \begin{bmatrix}  \ivec \\ \evec \end{bmatrix}_{\alpha_{N-1}}.
 \end{split}
\end{equation}
A similar explicit TT representation appears for the high-dimensional Laplace-\cite{khkaz-lap-2012}, and diffusion~\cite{kva-anidiff-2013} operators for high--dimensional PDEs.
Note also a related work on explicit tensor product representation of Fourier transform operator~\cite{dks-ttfft-2012,sav-rank1-2012}.

As noted in~\cite{ost-latensor-2009}, linear operations between vectors and matrices in tensor product formats can be computed efficiently in the same format.
In particular, the Hadamard product of vectors $[I(\x)]$ and $[p(\x)]$ can be computed efficiently in TT format with TT ranks of the product being the product of TT ranks of the terms~\cite{osel-tt-2011}.
Since multiplication by $[I(\x)]$ only doubles the TT ranks of $[p(\x)],$ the total complexity remains $\O(Nr^2).$
To compute the variance in~\eqref{eq:CME.stat}, we can use the formula 
\(
\Var{I}
  = \Mean{ I^2 } - (\Mean{I})^2,
\)
for which we need
\[
 \Mean{ I^2 } 
 = \sum_{\x\in\Omega} I^2(\x) p(\x)
 = | [ I(\x)^2 p(\x) ]_{\x\in\Omega} |_1
 = | [ I(\x) ]_{\x\in\Omega} \odot [ I(\x) ]_{\x\in\Omega} \odot \p |_1.
\]
Since each Hadamard multiplication with $[I(\x)]_{\x\in\Omega}$ doubles the TT ranks, we can apply them in order and then evaluate the sum, keeping total complexity to $\O(N r^2).$
However, we can suggest a more elegant explicit formula for the TT factorisation of tensor $[I(x)^2]_{\x\in\Omega}$ with all TT ranks equal to three.
From~\eqref{eq:I.cp} we obtain
\begin{equation}\label{eq:I2.cp}
 \begin{split}
   I(\x)^2
    & = \left(
        \sum_{n=1}^N \Indicator{x_n = \mathrm{i}}
        \right)^2
      = \underbrace{\sum_{1\leq n\leq N} \left(\Indicator{x_n = \mathrm{i}} \right)^2}_{I(\x)}
       + 2 \underbrace{\sum_{1\leq m < n \leq N} \Indicator{x_m = \mathrm{i}}\Indicator{x_n = \mathrm{i}}}_{B(\x)}.
 \end{split}
\end{equation}
Since $\Indicator{}^2 = \Indicator{},$ the first term equals $I(\x)$ and the TT decomposition is given by~\eqref{eq:I.tt}.
The second term is a sum of $\tfrac12 N(N-1)$ rank-one terms 
\(
\evec \otimes\cdots\otimes 
\evec \otimes \ivec \otimes 
\evec \otimes\cdots\otimes
\evec \otimes \ivec \otimes 
\evec \otimes\cdots\otimes
\evec,
\)
where $\ivec$'s appear in positions $m$ and $n,$ $n>m.$
Collecting linearly independent terms in each variable similarly to~\cite{khkaz-lap-2012}, we arrive at
a TT representation of ranks three:
\[
 [B(\x)]_{\x\in\Omega}
    = \sum_{\alpha_1,\ldots,\alpha_{N-1}=1}^{3} 
       \begin{bmatrix} \evec & \ivec & \zero \end{bmatrix}_{\alpha_1} 
        \otimes
        \begin{bmatrix} \evec & \ivec & \zero \\ \zero & \evec & \ivec \\ \zero & \zero & \evec \end{bmatrix}_{\alpha_1,\alpha_2} 
        \otimes\cdots\otimes 
        \begin{bmatrix} \evec & \ivec & \zero \\ \zero & \evec & \ivec \\ \zero & \zero & \evec \end{bmatrix}_{\alpha_{N-2},\alpha_{N-1}} 
         \otimes
        \begin{bmatrix} \zero \\ \ivec \\ \evec \end{bmatrix}_{\alpha_{N-1}}.
\]
Extending the TT representation~\eqref{eq:I.tt} to TT rank three by zero-padding the first and the last TT core allows us to represent
\(
I(\x)^2 = I(\x)+2B(\x)
\)
as a TT decomposition of TT ranks three:
\begin{equation}\label{eq:I2.tt}
 [I(\x)^2]_{\x\in\Omega}
    = \sum_{\alpha_1,\ldots,\alpha_{N-1}=1}^{3} 
       \begin{bmatrix} \evec & \ivec & \zero \end{bmatrix}_{\alpha_1} 
        \otimes
        \begin{bmatrix} \evec & \ivec & \zero \\ \zero & \evec & \ivec \\ \zero & \zero & \evec \end{bmatrix}_{\alpha_1,\alpha_2} 
        \otimes\cdots\otimes 
        \begin{bmatrix} \evec & \ivec & \zero \\ \zero & \evec & \ivec \\ \zero & \zero & \evec \end{bmatrix}_{\alpha_{N-2},\alpha_{N-1}} 
         \otimes
        \begin{bmatrix} \ivec \\ 2\ivec+\evec \\ 2\evec \end{bmatrix}_{\alpha_{N-1}}.
\end{equation}

In addition to the above, we may want to calculate so-called \emph{exceedance probabilities}
\begin{equation}\label{eq:exceed}
 \Prob( I(t) \geq I_\star  )
  = \sum_{\x\in\Omega} p(\x,t) \Indicator{I(\x) \geq I_\star}
  = \langle \p(t), [\Indicator{I(\x) \geq I_\star}]_{\x\in\Omega} \rangle,
\end{equation}
with some critical threshold $I_\star,$ e.g. related to a hospital capacity.
Similar to previous examples, we can evaluate this sum efficiently if we construct a TT representation for
\(
[ \Indicator{I(\x)\geq I_\star} ]_{\x\in\Omega}.
\)
Since 
\(
\Indicator{I(\x)\geq I_\star} = \sum_{I=I_\star}^N \Indicator{I(\x)=I},
\)
we start by constructing TT representations for
\(
[ \Indicator{I(\x) = I} ]_{\x\in\Omega}
\)
first.
The states $\x=(x_1,x_2,\ldots,x_N)^T$ with $I(\x)=I$ are such that exactly $I$ nodes are infected, and other $N-I$ nodes are not.
Extending the technique~\cite{khkaz-lap-2012} used to derive~\eqref{eq:I.tt} and~\eqref{eq:I2.tt}, the indicators can be shown to have the following TT representations
\begin{equation}\label{eq:I=i.tt}
 \begin{split}
  [\Indicator{I(\x)=I}]_{\x\in\Omega} 
     = \sum_{\alpha_1,\ldots,\alpha_{N-1}=1}^{I+1} 
 & \mathbf{u}^{(1)}_{\alpha_1} \otimes \cdots \otimes
   \mathbf{u}^{(n)}_{\alpha_{n-1},\alpha_n} 
   \otimes \cdots \otimes
   \mathbf{u}^{(N)}_{\alpha_{N-1}},
   \\
  \mathbf{u}^{(1)} 
    = \begin{bmatrix} \evec-\ivec & \ivec & \zero & \cdots & \zero \end{bmatrix},
 \quad
  \mathbf{u}^{(n)} 
    & = \begin{bmatrix} 
         \evec-\ivec & \ivec       & \zero    & \cdots      & \zero \\
         \zero       & \evec-\ivec & \ivec    & \ddots       & \vdots\\
         \vdots      & \ddots      & \ddots   & \ddots      &  \zero \\
         \vdots      &             & \ddots   & \evec-\ivec & \ivec \\
        \zero       &  \cdots      &   \cdots &    \zero    & \evec-\ivec
      \end{bmatrix},
  \quad
  \mathbf{u}^{(N)} 
    = \begin{bmatrix} \zero \\ \vdots \\ \zero \\ \ivec \\ \evec-\ivec \end{bmatrix},
 \end{split}
\end{equation}
with TT cores for $n=2,\ldots,N-1$ being the same $(I+1) \times 3 \times (I+1)$ array, all $(I+1)\times(I+1)$ slices of which are two-diagonal Toeplitz matrices.
Note that all TT ranks of this decomposition are equal to $I+1.$
For $I(\x)>N/2$ it is more convenient to `flip' the variables and count how many people are not infected, which leads to the following decomposition
\begin{equation}\label{eq:I=n-h.tt}
 \begin{split}
  [\Indicator{I(\x)=N-H}]_{\x\in\Omega} 
     = \sum_{\alpha_1,\ldots,\alpha_{N-1}=1}^{H+1} 
& \mathbf{v}^{(1)}_{\alpha_1} \otimes \cdots \otimes
  \mathbf{v}^{(n)}_{\alpha_{n-1},\alpha_n} 
   \otimes \cdots \otimes
  \mathbf{v}^{(N)}_{\alpha_{N-1}},
   \\
  \mathbf{v}^{(1)} 
    = \begin{bmatrix} \ivec & \evec-\ivec & \zero & \cdots & \zero \end{bmatrix},
 \quad
  \mathbf{v}^{(n)} 
    & = \begin{bmatrix} 
         \ivec       & \evec-\ivec & \zero       & \cdots  & \zero \\
         \zero       &       \ivec & \evec-\ivec & \ddots   & \vdots\\
         \vdots      &  \ddots     & \ddots      & \ddots  &  \zero \\
         \vdots      &             & \ddots      & \ivec   & \evec-\ivec \\
        \zero       &  \cdots      & \cdots      & \zero   & \ivec
      \end{bmatrix},
  \quad
  \mathbf{v}^{(N)} 
    = \begin{bmatrix} \zero \\ \vdots \\ \zero \\ \evec-\ivec \\ \ivec \end{bmatrix},
 \end{split}
\end{equation}
with all TT ranks are equal to $H+1=N-I+1.$
Summing the above equation for $H=0,\ldots,N-I_\star,$ we obtain the TT representation for the vector needed for computing the exceedance probability
\begin{equation}\label{eq:I>=I.tt}
 \begin{split}
  [\Indicator{I(\x)\geq I_\star}]_{\x\in\Omega} 
     = \sum_{\alpha_1,\ldots,\alpha_{N-1}=1}^{N-I_\star+1} 
& \mathbf{w}^{(1)}_{\alpha_1} \otimes \cdots \otimes
  \mathbf{w}^{(n)}_{\alpha_{n-1},\alpha_n} 
   \otimes \cdots \otimes
  \mathbf{w}^{(N)}_{\alpha_{N-1}},
   \\
  \mathbf{w}^{(1)} 
    = \begin{bmatrix} \ivec & \evec-\ivec & \zero & \cdots & \zero \end{bmatrix},
 \quad
  \mathbf{w}^{(n)} 
    & = \begin{bmatrix} 
         \ivec       & \evec-\ivec & \zero       & \cdots  & \zero \\
         \zero       &       \ivec & \evec-\ivec & \ddots  & \vdots\\
         \vdots      &   \ddots    & \ddots      & \ddots  &  \zero \\
         \vdots      &             & \ddots      & \ivec   & \evec-\ivec \\
        \zero       &  \cdots      & \cdots      & \zero   & \ivec
      \end{bmatrix},
  \quad
  \mathbf{w}^{(N)} 
    = \begin{bmatrix} \evec \\ \vdots \\ \evec \\ \evec \\ \ivec \end{bmatrix}.
 \end{split}
\end{equation}
Implementing this formula allows us to compute~\eqref{eq:exceed} in $\O(N r^2 (N-I_\star)^2)$ operations,
where $r$ is the largest TT rank of $\p.$

Finally, for a general observable that can be realised by an expectation 
\begin{equation}\label{eq:obs}
 q(t) 
  = \Mean{ Q(\x) } 
  = \sum_{\x\in\Omega} Q(\x) \p(\x,t)  
  = \langle \mathbf{Q}, \p(t) \rangle,
\end{equation}
we can compute a TT approximation of the vector 
\(
[Q(\x)]_{\x\in\Omega}
\)
by using TT cross interpolation methods~\cite{ot-ttcross-2010,so-dmrgi-2011proc,sav-qott-2014,ds-parcross-2020}.
These methods sample the function $Q(\x)$ typically at $\O(Nr^2)$ adaptively chosen states $\x\in\Omega$, followed by $\O(Nr^3)$ other floating point operations in linear algebra.

Note that observables are evaluated as a post-processing step after solving the master equation~\eqref{eq:CME}.
This is in contrast to using SSA, where the desired observations have to be stated in advance.

\section{Results}

The proposed method and necessary tensor product algorithms are implemented by authors in Matlab.
The SSA algorithm is implemented by authors in Matlab.
Where possible, the accuracy of results obtained by numerical methods is verified against analytic solutions, which were obtained as follows.
First, for a given network of contacts, the Markov chain transition graphs (such as the one in~Fig.~\ref{fig:CME}) were constructed and the ODEs~\eqref{eq:cme} were written using Julia language.
After that, the analytic solutions for the ODEs were obtained using SageMath software package, which runs Maxima computer algebra system as a backend.
The computations were performed in Matlab 2020b on an Intel Xeon E5-2640 v4 CPU with 2.40 GHz.

The codes are publicly available from
\begin{itemize}
 \item \href{https://github.com/savostyanov/ttsir}{github.com/savostyanov/ttsir}.
\end{itemize}

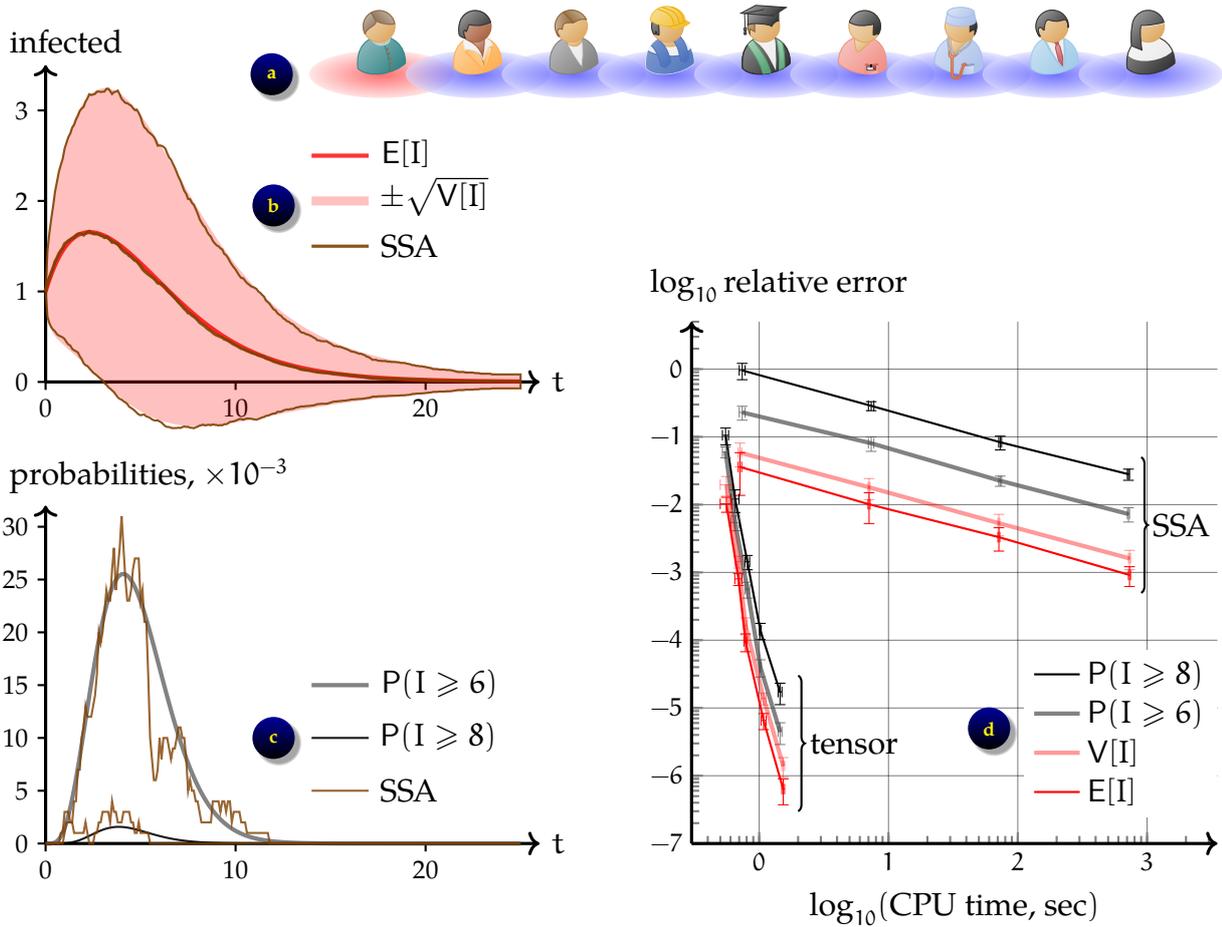
\begin{figure}[t]
   \beginpgfgraphicnamed{ttsir-pic2}  
   \begin{tikzpicture}[x=\linewidth,y=\linewidth]
    \node[lab] at(.175,.6){\textbf{a}};
    \begin{scope}[shift={(.26,.6)},x=1em,y=1em]
      \draw[Is]( 0,0)coordinate(p)ellipse(2.25em and .75em);\node[charlie,above,minimum size=1.5em,opacity=1] at (p){};
      \draw[Ss]( 3,0)coordinate(p)ellipse(2.25em and .75em);\node[alice,above,minimum size=1.5em,opacity=1] at (p){};
      \draw[Ss]( 6,0)coordinate(p)ellipse(2.25em and .75em);\node[bob,above,minimum size=1.5em,opacity=1] at (p){};
      \draw[Ss]( 9,0)coordinate(p)ellipse(2.25em and .75em);\node[builder,above,minimum size=1.5em,opacity=1] at (p){};
      \draw[Ss](12,0)coordinate(p)ellipse(2.25em and .75em);\node[graduate,above,minimum size=1.5em,opacity=1] at (p){};
      \draw[Ss](15,0)coordinate(p)ellipse(2.25em and .75em);\node[nurse,mirrored,above,minimum size=1.5em,opacity=1] at (p){};
      \draw[Ss](18,0)coordinate(p)ellipse(2.25em and .75em);\node[physician,mirrored,above,minimum size=1.5em,opacity=1] at (p){};
      \draw[Ss](21,0)coordinate(p)ellipse(2.25em and .75em);\node[dave,above,minimum size=1.5em,opacity=1] at (p){};
      \draw[Ss](24,0)coordinate(p)ellipse(2.25em and .75em);\node[nun,above,minimum size=1.5em,opacity=1] at (p){};
    \end{scope} 

      \begin{scope}[shift={(.0,.36)},x=2.5mm,y=12mm]
       \node[lab] at(12,1.95){\textbf{b}};
       \draw[->,very thick] (0,0) -- (26,0) node[right] {$t$};
       \draw[->,very thick] (0,0) -- (0,3.5) node[above right,xshift=-1.5em] {infected};
       \foreach \x in {0,10,20}{\draw[thick] (\x,0) -- (\x,-3pt) node[below,tick] {\x};}
       \foreach \y in {0,1,2,3}{\draw[thick] (0,\y) -- (-3pt,\y) node[left,tick] {\y};}
       \draw[\Icol,ultra thick,opacity=.8]
            plot file{./dat/sir/sir_chain09_mean00.dat}
            (14,2.5) --++(3,0) node[right,text opacity=1,black]{$\Mean{I}$};
       \fill[\Icol,opacity=.25]
            plot file{./dat/sir/sir_chain09_pmsd00.dat}
            (14,1.95) rectangle++(3,.1)node[right,text opacity=1,black]{$\pm\sqrt{\Var{I}}$};
       \draw[SSA,very thick,opacity=.9]  
            plot file{./dat/sir/ssa_chain09_imean03.dat}
            (14,1.5) --++(3,0)node[right,text opacity=1,black]{SSA};
       \draw[SSA,thick,opacity=.9] 
            plot file{./dat/sir/ssa_chain09_ipmsd03.dat};
      \end{scope} 
    
     \begin{scope}[shift={(.0,.0)}, x=2.5mm,y=1.4mm]
       \node[lab] at(12,10){\textbf{c}};
       \draw[->,very thick] (0,0) -- (26,0) node[right] {$t$};
       \draw[->,very thick] (0,0) -- (0,32) node[above right,xshift=-1.5em] {probabilities, $\times 10^{-3}$};
       \foreach \x in {0,10,20}{\draw[thick] (\x,0) -- (\x,-3pt) node[below,tick] {\x};}
       \foreach \y in {0,5,10,15,20,25,30}{\draw[thick] (0,\y) -- (-3pt,\y) node[left,tick] {\y};}
       \draw[black!70,ultra thick,opacity=.7] 
            plot file{./dat/sir/sir_chain09_igeq6_prob00_scaled.dat}
            (14,15)--++(3,0)node[right,text opacity=1,black]{$\Prob(I\geq6)$};
       \draw[black,thick,opacity=.9] 
             plot file{./dat/sir/sir_chain09_igeq8_prob00_scaled.dat} 
            (14,10)--++(3,0)node[right,text opacity=1,black]{$\Prob(I\geq8)$};
       \draw[SSA,thick,opacity=.8]  
            plot file{./dat/sir/ssa_chain09_igeq6_prob03_scaled.dat}
            (14,5)--++(3,0)node[right,text opacity=1,black]{SSA};
       \draw[SSA,thick,opacity=.8]  
             plot file{./dat/sir/ssa_chain09_igeq8_prob03_scaled.dat}
            ;
     \end{scope}
     
     \begin{scope}[shift={(.5,.0)}]
      \begin{axis}[width=.5\textwidth,height=.5\textwidth,%
        xmode=log, ymode=log,
        xlabel={$\log_{10}(\text{CPU time, sec})$},
        ylabel={$\log_{10}\text{relative error}$},
        xmin=3e-1,xmax=3.5e3,
        ymax=5e0,ymin=1e-7,
        every axis plot/.append style={
                        error bars/y dir=both,
                        error bars/y explicit,
                        error bars/x dir=both,
                        error bars/x explicit
                        },
        legend style={anchor=south east,at={(1,.05)}},
      ] 
  \addplot+[] coordinates{
( 0.73902, 9.566e-01) +- (0.03850 , 2.595e-01 )
( 7.35088, 2.870e-01) +- (0.354129, 4.360e-02 )
( 73.0673, 8.335e-02) +- (1.90667 , 1.914e-02 )
( 718.259, 2.815e-02) +- (14.3036 , 5.281e-03 )
  };
  \addplot+[] coordinates{
( 0.73902, 2.297e-01 ) +- (0.03850 , 5.330e-02)
( 7.35088, 8.048e-02 ) +- (0.354129, 1.932e-02)
( 73.0673, 2.251e-02 ) +- (1.90667 , 3.860e-03)
( 718.259, 7.268e-03 ) +- (14.3036 , 1.727e-03)
  };
  \addplot+[] coordinates{
( 0.70983, 5.861e-02) +- (0.022050, 2.243e-02)
( 7.07659, 1.804e-02) +- (0.129318, 6.210e-03)
( 71.4063, 5.328e-03) +- (1.22912 , 1.862e-03)
( 731.786, 1.611e-03) +- (14.0976 , 5.090e-04)
  };
  \addplot+[] coordinates{
( 0.70983, 3.605e-02) +- (0.022050, 2.235e-02)
( 7.07659, 1.013e-02) +- (0.129318, 4.872e-03)
( 71.4063, 3.315e-03) +- (1.22912 , 1.253e-03)
( 731.786, 9.173e-04) +- (14.0976 , 2.974e-04)
  };


  \addplot+[] coordinates{
(0.549509 , 1.055e-01) +- (0.0302754, 2.950e-02)
(0.651331 , 1.209e-02) +- (0.0459733, 4.431e-03)
(0.809719 , 1.437e-03) +- (0.0343232, 3.411e-04)
(1.02109  , 1.396e-04) +- (0.0161838, 3.753e-05)
(1.44984  , 1.720e-05) +- (0.049612 , 6.001e-06)
  };
 \addplot+[] coordinates{
(0.549509 , 6.010e-02) +- (0.0302754, 1.109e-02)
(0.651331 , 5.458e-03) +- (0.0459733, 1.340e-03)
(0.809719 , 5.660e-04) +- (0.0343232, 1.489e-04)
(1.02109  , 4.000e-05) +- (0.0161838, 1.141e-05)
(1.44984  , 4.471e-06) +- (0.049612 , 1.582e-06)
  };
  \addplot+[] coordinates{
(0.556909, 1.968e-02) +- (0.0573936, 6.322e-03)
(0.69138 , 1.445e-03) +- (0.0436877, 2.859e-04)
(0.783876, 1.665e-04) +- (0.0230046, 4.660e-05)
(1.08172 , 1.386e-05) +- (0.0482557, 2.750e-06)
(1.53327 , 1.400e-06) +- (0.0384052, 4.529e-07)
  };
  \addplot+[] coordinates{
(0.556909, 1.028e-02) +- (0.0573936, 2.568e-03)
(0.69138 , 8.042e-04) +- (0.0436877, 1.628e-04)
(0.783876, 9.574e-05) +- (0.0230046, 2.816e-05)
(1.08172 , 6.537e-06) +- (0.0482557, 1.748e-06)
(1.53327 , 6.358e-07) +- (0.0384052, 2.631e-07)
  };
 
  \legend{$\Prob(I\geq 8)$, $\Prob(I\geq 6)$, $\Var{I}$, $\Mean{I}$}
  
  \draw[thick,decoration={brace},decorate] (axis cs:9e2,5e-2) -- (axis cs:9e2,5e-4) node[midway,right]{SSA};
  \draw[thick,decoration={brace},decorate] (axis cs:2e0,3e-5) -- (axis cs:2e0,3e-7) node[midway,right]{tensor};
  \node[lab] at(axis cs:6e1,5e-6){\textbf{d}};
   \end{axis}
  \end{scope}
\end{tikzpicture}
\endpgfgraphicnamed
 \caption{SIR epidemic on a linear chain of $N=9$ people, shown for $\beta=1$ and $\gamma=0.3$:
            (a) the network in its initial state;
            (b) mean and variance of the number of infected computed analytically 
                and simulated using SSA with $\Nssa = 10^3$ sampled trajectories;
            (c) exceedance probabilities computed analytically and simulated using SSA with $\Nssa = 10^3$ sampled trajectories;
            (d) relative error of mean, variance, and exceedance probabilities, computed by SSA and tensor product approach.
                 }
  \label{fig:chain}
\end{figure}

\begin{figure}[t]
  \beginpgfgraphicnamed{ttsir-pic3}  
   \begin{tikzpicture}[x=\linewidth,y=\linewidth]
    \begin{scope}[shift={(.3,.45)}]
     \tikzstyle{city}=[draw=none,fill=orange]
     \tikzstyle{road}=[double=orange,double distance=.5em,opacity=.5,postaction={draw=white,very thick,dashed}]
     \node[anchor=south west,inner sep=0](map)at(0,0){\includegraphics[width=.6\textwidth]{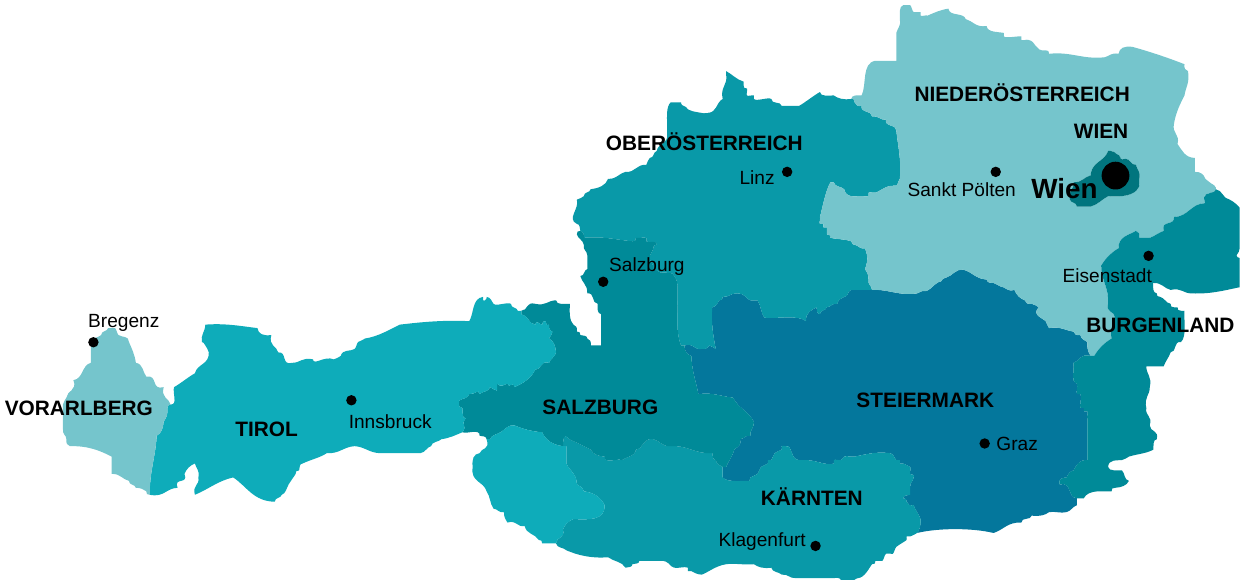}};
     \begin{scope}[x={(map.south east)},y={(map.north west)}]
      \fill[white,opacity=.6] (0,0)rectangle(1,1);
      \node[lab] at(.35,.7){\textbf{a}};

      \coordinate(vorarlberg)at(.1,.25);
      \coordinate(tirol)at(.3,.33);
      \coordinate(salzburg)at(.5,.6);
      \coordinate(karnten)at(.65,.2);
      \coordinate(oberosterreich)at(.65,.75);
      \coordinate(niederosterreich)at(.8,.75);
      \coordinate(steiermark)at(.75,.35);
      \coordinate(burgenland)at(.91,.4);
      \coordinate(wien)at(.95,.7);

      \draw[road]
      (vorarlberg)
            to[out=10,in=180] (tirol)
            to[out=0,in=180] ++(.18,0)coordinate(x1)  to[out=0,in=225](salzburg)
            to[out=45,in=180] (oberosterreich)
            to[out=0,in=180] (niederosterreich)
            to[out=0,in=135] (wien)
      (salzburg)
            to[out=-45,in=135](karnten)
      (salzburg)
            to[out=-30,in=135](steiermark)
      (oberosterreich)
            to[out=-45,in=135](steiermark)
        (x1)to[out=0,in=135](karnten)
            to[out=-45,in=225](steiermark)
            to[out=45,in=90](niederosterreich)
      (steiermark)
            to[out=45,in=270](burgenland)
            to[out=90,in=-45](niederosterreich)
         ;
      \coordinate(p)at(vorarlberg);        \draw[Is](p)ellipse(2em and .6em); \node[charlie,above,minimum size=1.5em,opacity=1] at (p){};
      \coordinate(p)at(tirol);             \draw[Ss](p)ellipse(2em and .6em); \node[alice,above,minimum size=1.5em,opacity=1] at (p){};
      \coordinate(p)at(salzburg);          \draw[Ss](p)ellipse(2em and .6em); \node[bob,above,minimum size=1.5em,opacity=1] at (p){};
      \coordinate(p)at(karnten);           \draw[Ss](p)ellipse(2em and .6em); \node[dave,above,minimum size=1.5em,opacity=1] at (p){};
      \coordinate(p)at(oberosterreich);    \draw[Ss](p)ellipse(2em and .6em); \node[charlie,mirrored,above,minimum size=1.5em,opacity=1] at (p){};
      \coordinate(p)at(steiermark);        \draw[Ss](p)ellipse(2em and .6em); \node[alice,mirrored,above,minimum size=1.5em,opacity=1] at (p){};
      \coordinate(p)at(niederosterreich);  \draw[Ss](p)ellipse(2em and .6em); \node[bob,mirrored,above,minimum size=1.5em,opacity=1] at (p){};
      \coordinate(p)at(burgenland);        \draw[Ss](p)ellipse(2em and .6em); \node[dave,mirrored,above,minimum size=1.5em,opacity=1] at (p){};
      \coordinate(p)at(wien);              \draw[Ss](p)ellipse(2em and .6em); \node[alice,mirrored,above,minimum size=1.5em,opacity=1] at (p){};
     \end{scope}
    \end{scope}

    \begin{scope}[shift={(.0,.37)},x=2.5mm,y=8.5mm]
     \node[lab] at(14,1.5){\textbf{b}};
     \draw[->,very thick] (0,0) -- (26,0) node[right] {$t$};
     \draw[->,very thick] (0,0) -- (0,6.3) node[above right,xshift=-1.5em] {infected};
     \foreach \x in {0,10,20}{\draw[thick] (\x,0) -- (\x,-3pt) node[below,tick] {\x};}
     \foreach \y in {0,1,2,3,4,5,6}{\draw[thick] (0,\y) -- (-3pt,\y) node[left,tick] {\y};}
     \draw[\Icol,ultra thick,opacity=.8]
            plot file{./dat/sir/sir_a09_mean00.dat}
            (16,2.2) --++(3,0) node[right,text opacity=1,black]{$\Mean{I}$};
     \fill[\Icol,opacity=.25]
            plot file{./dat/sir/sir_a09_pmsd00.dat}
            (16,1.5) rectangle++(3,.15)node[right,text opacity=1,black]{$\pm\sqrt{\Var{I}}$};
     \draw[SSA,very thick,opacity=.9]  
            plot file{./dat/sir/ssa_a09_imean03.dat}
            (16,0.9) --++(3,0)node[right,text opacity=1,black]{SSA};
     \draw[SSA,thick,opacity=.9] 
            plot file{./dat/sir/ssa_a09_ipmsd03.dat};
    \end{scope} 
    
   \begin{scope}[shift={(.0,.0)}, x=2.5mm,y=.18mm]
     \node[lab] at(14,70){\textbf{c}};
     \draw[->,very thick] (0,0) -- (26,0) node[right] {$t$};
     \draw[->,very thick] (0,0) -- (0,250) node[above right,xshift=-1.5em] {probabilities, $\times 10^{-3}$};
     \foreach \x in {0,10,20}{\draw[thick] (\x,0) -- (\x,-3pt) node[below,tick] {\x};}
     \foreach \y in {0,50,100,150,200}{\draw[thick] (0,\y) -- (-3pt,\y) node[left,tick] {\y};}
     \draw[black!70,ultra thick,opacity=.7] 
            plot file{./dat/sir/sir_a09_igeq6_prob00_scaled.dat}
            (16,100)--++(3,0)node[right,text opacity=1,black]{$\Prob(I\geq6)$};
     \draw[black,thick,opacity=.9] 
             plot file{./dat/sir/sir_a09_igeq8_prob00_scaled.dat} 
            (16, 70)--++(3,0)node[right,text opacity=1,black]{$\Prob(I\geq8)$};
     \draw[SSA,thick,opacity=.8]
            plot file{./dat/sir/ssa_a09_igeq6_prob03_scaled.dat}
            (16, 40)--++(3,0)node[right,text opacity=1,black]{SSA};
     \draw[SSA,thick,opacity=.8]
             plot file{./dat/sir/ssa_a09_igeq8_prob03_scaled.dat}
            ;
   \end{scope}
   
   \begin{scope}[shift={(.5,.0)}]
    \begin{axis}[width=.5\textwidth,height=.5\textwidth,%
        xmode=log, ymode=log,
        xlabel={$\log_{10}(\text{CPU time, sec})$},
        ylabel={$\log_{10}\text{relative error}$},
        xmin=3e-1,xmax=3.5e3,
        ymax=5e0,ymin=1e-7,
        every axis plot/.append style={
                        error bars/y dir=both,
                        error bars/y explicit,
                        error bars/x dir=both,
                        error bars/x explicit
                        },
        legend style={anchor=south east,at={(1,.05)}},
      ] 
  \addplot+[] coordinates{
(1.10488, 1.648e-01) +- (0.040773, 3.343e-02 )
(10.9564, 6.454e-02) +- (0.192114, 2.315e-02 )
(109.469, 1.903e-02) +- (2.20558 , 6.588e-03 )
(1112.45, 5.681e-03) +- (25.3022 , 1.549e-03 )
  };
  \addplot+[] coordinates{
(1.10488, 6.837e-02) +- (0.040773, 2.397e-02 )
(10.9564, 2.204e-02) +- (0.192114, 6.265e-03 )
(109.469, 7.707e-03) +- (2.20558 , 2.408e-03 )
(1112.45, 2.305e-03) +- (25.3022 , 5.901e-04 )
  };
  \addplot+[] coordinates{
(1.06685, 3.090e-02) +- (0.0378769, 7.943e-03)
(10.2781, 1.044e-02) +- (0.299978 , 2.499e-03)
(103.634, 3.379e-03) +- (2.86026  , 1.107e-03)
(1075.1 , 1.165e-03) +- (18.0549  , 3.339e-04)
  };
  \addplot+[] coordinates{
(1.06685, 2.825e-02) +- (0.0378769, 8.305e-03)
(10.2781, 8.300e-03) +- (0.299978 , 3.915e-03)
(103.634, 2.707e-03) +- (2.86026  , 1.132e-03)
(1075.1 , 8.952e-04) +- (18.0549  , 3.182e-04)
  };
  
  \addplot+[] coordinates{
(0.626598, 1.698e-02) +- (0.0291144, 4.288e-03)
(0.799166, 2.219e-03) +- (0.0221837, 5.189e-04)
(1.13337 , 2.445e-04) +- (0.0261624, 7.664e-05)
(1.865   , 2.993e-05) +- (0.0596819, 9.969e-06)
(3.62998 , 3.232e-06) +- (0.0896744, 5.078e-07)
  };
 \addplot+[] coordinates{
(0.626598, 1.611e-02) +- (0.0291144, 3.197e-03)
(0.799166, 2.171e-03) +- (0.0221837, 4.104e-04)
(1.13337 , 1.747e-04) +- (0.0261624, 5.438e-05)
(1.865   , 2.330e-05) +- (0.0596819, 6.912e-06)
(3.62998 , 3.074e-06) +- (0.0896744, 5.987e-07)
  };
  \addplot+[] coordinates{
(0.64237,1.174e-02) +- (0.0256991,5.429e-03)
(0.82629,1.607e-03) +- (0.0660058,3.547e-04)
(1.15507,2.047e-04) +- (0.0363414,4.608e-05)
(1.83794,2.060e-05) +- (0.032209 ,3.429e-06)
(3.73907,3.315e-06) +- (0.103556 ,5.637e-07)
  };
  \addplot+[] coordinates{
(0.64237, 9.503e-03) +- (0.0256991, 4.643e-03)
(0.82629, 1.083e-03) +- (0.0660058, 2.851e-04)
(1.15507, 1.093e-04) +- (0.0363414, 2.385e-05)
(1.83794, 9.350e-06) +- (0.032209 , 1.796e-06)
(3.73907, 1.557e-06) +- (0.103556 , 3.176e-07)
  };
 
  \legend{$\Prob(I\geq 8)$, $\Prob(I\geq 6)$, $\Var{I}$, $\Mean{I}$}
  
  \draw[thick,decoration={brace},decorate] (axis cs:1.4e3,1.4e-2) -- (axis cs:1.4e3,4e-4) node[midway,right](ssa){};
  \draw[thick,decoration={brace},decorate] (axis cs:5e0,8e-6) -- (axis cs:5e0,8e-7) node[midway,right]{tensor};
  \node[lab] at(axis cs:6e1,3e-6){\textbf{d}};
      
  \end{axis}
  \node[right] at(ssa){SSA};
 \end{scope}
\end{tikzpicture}
\endpgfgraphicnamed
 \caption{SIR epidemic on a road network in Austria, shown for $\beta=1$ and $\gamma=0.3$:
            (a) the network in its initial state;
            (b) mean and variance of the number of infected computed analytically 
                and simulated using SSA with $\Nssa = 10^3$ sampled trajectories;
            (c) exceedance probabilities computed analytically and simulated using SSA with $\Nssa = 10^3$ sampled trajectories;
            (d) relative error of mean, variance, and exceedance probabilities, computed by SSA and tensor product approach.
                 }
  \label{fig:austria}
\end{figure}

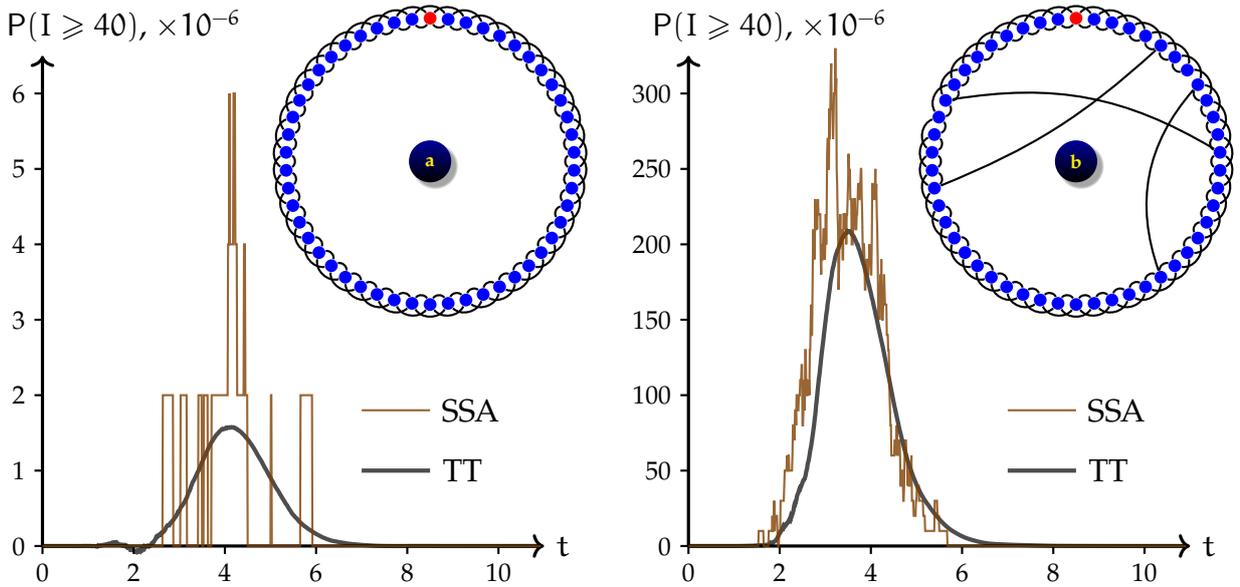
\begin{figure}[t]
   \beginpgfgraphicnamed{ttsir-pic4}  
   \begin{tikzpicture}[x=\linewidth,y=\linewidth]

    \begin{scope}[shift={(.3,.3)}]
     \node[lab] at(0,0){\textbf{a}};
     \pgfmathsetmacro{\N}{50}
     \foreach \n in {1,2,...,\N}{
      \pgfmathsetmacro{\a}{(360/\N)*(\n-1)+90}
      \ifnum \n=1
       \node[circle,fill=\Icol,inner sep=0,minimum width=.4em](\n) at({\a}:4.5em){};
      \else
       \node[circle,fill=\Scol,inner sep=0,minimum width=.4em](\n) at({\a}:4.5em){};
      \fi
     }
     \foreach \n in {1,2,...,\N}{
      \pgfmathsetmacro{\one}{int(mod(\n  ,\N)+1)}     \pgfmathsetmacro{\bendone}{60}
      \pgfmathsetmacro{\two}{int(mod(\n+1,\N)+1)}     \pgfmathsetmacro{\bendtwo}{60}
      \draw[thick] (\n) to[ bend left=\bendone] (\one);
      \draw[thick] (\n) to[bend right=\bendtwo] (\two);
     }
    \end{scope}
    \begin{scope}[shift={(.8,.3)}]
     \node[lab] at(0,0){\textbf{b}};
     \pgfmathsetmacro{\N}{50}
     \foreach \n in {1,2,...,\N}{
      \pgfmathsetmacro{\a}{(360/\N)*(\n-1)+90}
      \ifnum \n=1
       \node[circle,fill=\Icol,inner sep=0,minimum width=.4em](\n) at({\a}:4.5em){};
      \else
       \node[circle,fill=\Scol,inner sep=0,minimum width=.4em](\n) at({\a}:4.5em){};
      \fi
     }
     \foreach \n in {1,2,...,\N}{
      \pgfmathsetmacro{\one}{int(mod(\n  ,\N)+1)}     \pgfmathsetmacro{\bendone}{60}
      \pgfmathsetmacro{\two}{int(mod(\n+1,\N)+1)}     \pgfmathsetmacro{\bendtwo}{60}
      \ifnum \n=10
        \pgfmathsetmacro{\two}{39}  \pgfmathsetmacro{\bendtwo}{-20}
      \fi
      \ifnum \n=15
        \pgfmathsetmacro{\one}{46}  \pgfmathsetmacro{\bendone}{-10}
      \fi
      \ifnum \n=43
        \pgfmathsetmacro{\one}{31}  \pgfmathsetmacro{\bendone}{-30}
      \fi
      \draw[thick] (\n) to[ bend left=\bendone] (\one);
      \draw[thick] (\n) to[bend right=\bendtwo] (\two);
     }
    \end{scope}

   \begin{scope}[shift={(.0,.0)}, x=6mm,y=10mm]
     \draw[->,very thick] (0,0) -- (11,0) node[right] {$t$};
     \draw[->,very thick] (0,0) -- (0,6.5) node[above right,xshift=-1.5em] {$\Prob(I\geq40)$, $\times 10^{-6}$};
     \foreach \x in {0,2,4,6,8,10}{\draw[thick] (\x,0) -- (\x,-3pt) node[below,tick] {\x};}
     \foreach \y in {0,1,2,3,4,5,6}{\draw[thick] (0,\y) -- (-3pt,\y) node[left,tick] {\y};}
     \begin{scope}
      \clip (0,-1) rectangle (10.9,7);
      \draw[black,ultra thick,opacity=.7] 
            plot file{./dat/sir/tt_sw50p0_pge40_scaled.dat}
            (7,1.0)--++(1.5,0)node[right,text opacity=1,black]{TT};
      \draw[SSA,thick,opacity=.8]
            plot file{./dat/sir/ssa5e5_sw50p0_pge40_scaled.dat}
            (7,1.8)--++(1.5,0)node[right,text opacity=1,black]{SSA};
     \end{scope}
   \end{scope}
   \begin{scope}[shift={(.5,.0)}, x=6mm,y=.2mm]
     \draw[->,very thick] (0,0) -- (11,0) node[right] {$t$};
     \draw[->,very thick] (0,0) -- (0,325) node[above right,xshift=-1.5em] {$\Prob(I\geq40)$, $\times 10^{-6}$};
     \foreach \x in {0,2,4,6,8,10}{\draw[thick] (\x,0) -- (\x,-3pt) node[below,tick] {\x};}
     \foreach \y in {0,50,100,150,200,250,300}{\draw[thick] (0,\y) -- (-3pt,\y) node[left,tick] {\y};}
     \begin{scope}
      \clip (0,-50) rectangle (10.9,350);
      \draw[black,ultra thick,opacity=.7] 
            plot file{./dat/sir/tt7_sw50p3_pge40_scaled.dat}
            (7,50)--++(1.5,0)node[right,text opacity=1,black]{TT};
      \draw[SSA,thick,opacity=.8]
            plot file{./dat/sir/ssa_sw50p3_pge40_scaled.dat}
            (7,90)--++(1.5,0)node[right,text opacity=1,black]{SSA};
     \end{scope}
   \end{scope}
    
%
%
%
   \end{tikzpicture}
   \endpgfgraphicnamed
 \caption{SIR epidemic on small world networks with $N=50$ people, with probability of rare event $\Prob(I\geq40)$  computed using TT and SSA shown for $\beta=1$ and $\gamma=0.3$:
            (a) the small world network with short--range connections, where each person is linked with two neighbours left of them and two neighbours right of them on the circle;
            (b) the same network but with $3$ randomly selected nodes rewired to a random node.
                 }
  \label{fig:sw}
\end{figure}

\subsection{Linear chain network}
As a first experiment, we consider a linear network of $N=9$ people.
Out of $|\Omega|=3^N=19683$ network states only $1022=2(2^N-1)$ are accessible from the initial state.
This relatively modest scale of the problem makes it possible to write the ODEs~\eqref{eq:cme} and to solve them analytically using SageMath software, which took us about $7$ days of CPU time.
From the analytic expressions for the p.d.f. $\p_\star(t)$ we evaluated analytic expressions for observables~\eqref{eq:CME.stat} and~\eqref{eq:exceed} and used them as reference values.
The observables $q(t)$ obtained by numerical algorithms were compared with the reference values $q_\star(t)$ and the relative accuracy (relative error)  was measured as
\begin{equation}\label{eq:relerr}
 \text{relative accuracy} = \dfrac{\| q - q_\star \|_{L_2}}{\| q_\star \|_{L_2}},
 \qquad
 \|q(t)\|_{L_2}^2 = \int_0^\infty |q(t)|^2 \dt.
\end{equation}

In Fig.~\ref{fig:chain}, we show the relative errors~\eqref{eq:relerr} and CPU times of the TT and SSA methods for both the total mean number of infected individuals, and the occupancy probabilities for the linear chain.
It should be noted that the TT algorithms are parameterised by the error tolerance, which is used as a threshold for the relative error in the Frobenius norm for both the truncation of the TT decompositions and for stopping of the tAMEn algorithm.
In contrast, the SSA method is parameterised by the number of samples $\Nssa$.
Since these parameters don't match directly, we compare the CPU times of both methods.
For all quantities of interest, SSA converges with a $\O(\Nssa^{-1/2})$ rate as expected.
Although a seemingly modest number of samples may be sufficient to estimate mean population numbers, probabilities of rare events are much more difficult to estimate.
In particular, SSA gives a rather misleading information about the high occupancy probability even with tens thousands of samples, requiring hundreds of seconds of computing for this (relatively simple) example.
In contrast, the TT approach can compute the entire p.d.f. (and hence any derived statistics) with 4 accurate decimal digits in just a couple of seconds.

\subsection{Road network in Austria}

Now we test the methods on a network shown in Fig~\ref{fig:austria}(a), which illustrates the Austrian state adjacency map.
This network has $N=9$ nodes but more edges than the linear chain, resulting in better mixing and higher number of accessible network states.
Assume that the initial state is deterministic with the first node in the infected state and all other nodes in the susceptible state, the Markov chain has $4982$ accessible states for this network compared to $1022$ for the linear chain network of the same size.
Nevertheless, we were able to compute the analytic solution for this problem using SageMath and used it as a reference to benchmark the accuracy of numerical algorithms.

The accuracy of tAMEn and SSA algorithms is shown in Fig.~\ref{fig:austria}(d).
Similarly to the results in Fig.~\ref{fig:chain}(d), we see that SSA converges according to the central limit theorem law, whereas the TT method can achieve a faster rate.
Due to a more connected network, the exceedance probabilities are about ten times larger than those in the chain network, which makes it easier for the SSA algorithm to recover them.
However, as the geometry of this network is somewhat elongated in one direction, and matches (although not ideally) the linear geometry of the tensor train format, the tAMEn algorithm also performs well.
Overall, we can see that if two or more accurate digits are desired in observables, the tensor product approach is more attractive for this example

\subsection{Small world networks}
Lastly, we consider two small world networks, produced by the Watts--Strogatz algorithm, as shown in Figure~\ref{fig:sw}.
The first contains $N=50$ people, each connected to $2$ next and $2$ previous neighbours on a circular chain.
The second additionally has $3$ of its edges rewired to random vertices.
The epidemics starts with the first node in the infected state, all others in the susceptible state. 
We are interested in the probability that $I \geq 40$, i.e. $4/5$ of the population is infected at once.
For the first network,  $\Prob(I\geq40)$ reaches the level of $1.5 \cdot 10^{-6}$ at its peak, which makes it a rare event.
To accurately resolve this small value, we apply the tAMEn algorithm with the approximation threshold of $10^{-9}.$
The TT ranks of the p.d.f. $\p(t)$ reach the value of $401$, and the computation takes about 5 hours of CPU time.
We then applied the SSA method with $5 \cdot 10^5$ trajectories, which requires approximately the same CPU time, and compared the results in Figure~\ref{fig:sw}(a).
We see that only a tiny fraction of SSA trajectories has hit the event of interest, resulting in a rather inaccurate estimate of the probability.
The TT method was able to produce more accurate and smooth estimate of the probability.
We note that the probability recovered by the TT approach has a numerical artefact at $t\approx2$ where it became negative with the magnitude of $10^{-7},$ which is caused by the approximation error.
This suggests that our choice of the error threshold is reasonable, as more aggressive compression may destroy the structure of the probability of interest, while a more accurate approximation would result in larger TT ranks and CPU time.
To reach a similar accuracy with SSA we would have to increase the number of trajectories to $10^7$ that would require more than $120$ hours of computing, while the tAMEn algorithm recovers the whole p.d.f. in just 5 hours.

The rewired network has more long--range connections that facilitate the propagation of the infection, and makes it much more probable for a large number of people to be infected at once.
For instance, for this network $\Prob(I\geq40)$ peaks at about $2\cdot10^{-4},$ making it easier for SSA to discover this event.
With only $10^5$ trajectories, SSA yields a reasonable estimate in $1.3$ hours.
For the TT method we can take the approximation threshold of $10^{-7},$  observe the maximal TT rank of the p.d.f. to be $337$, and recover $\p(t)$ using $2.5$  hours of CPU time.
The results are compared in~Fig.~\ref{fig:sw}(b), where we again can see that the estimate obtained by the TT algorithm is more accurate.

We see that the TT method  is beneficial for rare event simulations requiring high overall accuracy and/or weakly correlated systems, since the TT decomposition converges rapidly when the correlations are local.




\section{Discussion and conclusion}
We have demonstrated numerically that the TT approximation of the probability distribution function converges rapidly for stochastic population models on networks with local connections,
  with only a polynomial scaling in the number of individuals,
  and a poly-logarithmic scaling in the error.
This allows one to compute any statistics of such models more accurately than using the stochastic simulation.
In particular, we have managed to compute probabilities of high infectivity events of the order of $10^{-6}$.
Thus, tensor methods can be recommended for models of moderate size, local connectivity, and/or if rare event statistics are of interest.

Two limitations of the proposed approach are very high dimensions and long--range connections in the network.
The TT approximations are based on the singular value decomposition,
   which is known to produce an optimal approximation in the 2-norm.
However, observables depend linearly on probabilities of the network states, and the perturbation in observables computed with a CME model can be bounded by a 1-norm error in the p.d.f.
In high dimensions, the equivalence constant between 1- and 2-norms can be large, which requires one to use a very small truncation threshold in the 2-norm.
Since
\(
\|\p\|_1 = \|\sqrt{\p}\|_2^2,
\)
this problem can be solved by reformulating the CME into a nonlinear ODE for $\sqrt{p(\x,t)}.$
However, this nonlinear ODE features the reciprocals $1/p(\x,t),$ preventing calculations whenever zero probabilities are present, for example, when the initial state is deterministic.

Long--range connections in the network inflate the ranks of the TT decomposition and slow the method down.
This drawback may be curable by using tree tensor networks with the topology adapted to the given network \cite{legeza-graph-opt-2011,ballani-adapttree-2014,bebe-maca-2014}.
This is a subject of future work.

Finally, we should note that SSA algorithm is embarrassingly parallel, because all trajectories can be sampled independently.
To be competitive with SSA, tensor product algorithms also have to perform well on distributed--memory high--performance computing platforms. 
Recent progress in this area includes~\cite{solomonik-cyclops-2014,grasedyck-par-cross-2015,sgldcj-pTDVP-2020,ds-parcross-2020}.

\newpage

\begin{thebibliography}{10}

\bibitem{Ammar-cme-2011}
A.~Ammar, E.~Cueto, and F.~Chinesta.
\newblock \href{https://doi.org/10.1002/cnm.2476}{Reduction of the chemical
  master equation for gene regulatory networks using proper generalized
  decompositions}.
\newblock {\em Int. J. Numer. Meth. Biomed. Engng.}, 28\penalty0 (9):\penalty0
  960--973, 2012.

\bibitem{Anderson-tau-leaping-2011}
David~F. Anderson, Arnab Ganguly, and Thomas~G. Kurtz.
\newblock \href{https://doi.org/10.1214/10-AAP756}{Error analysis of {Tau-Leap}
  simulation methods}.
\newblock {\em The Annals of Applied Probability}, 21\penalty0 (6):\penalty0
  2226--2262, 2011.

\bibitem{AndersonHigham-MLCME-2012}
David~F. Anderson and Desmond~J. Higham.
\newblock \href{https://doi.org/10.1137/110840546}{Multilevel {Monte Carlo} for
  continuous time {Markov} chains, with applications in biochemical kinetics}.
\newblock {\em Multiscale Modeling \& Simulation}, 10\penalty0 (1):\penalty0
  146--179, 2012.

\bibitem{ballani-adapttree-2014}
J.~Ballani and L.~Grasedyck.
\newblock \href{https://doi.org/10.1137/130926328}{Tree adaptive approximation
  in the hierarchical tensor format}.
\newblock {\em SIAM J. Sci. Comput.}, 36\penalty0 (4):\penalty0 A1415--A1431,
  2014.

\bibitem{lars-review-2014}
Jonas Ballani, Lars Grasedyck, and Melanie Kluge.
\newblock \href{https://doi.org/10.1007/978-3-319-08159-5_10}{A review on
  adaptive low-rank approximation techniques in the hierarchical tensor
  format}.
\newblock In {\em Extraction of Quantifiable Information from Complex Systems},
  volume 102 of {\em Lecture Notes in Computational Science and Engineering},
  Springer, 2014, pages 195--210.

\bibitem{legeza-graph-opt-2011}
G.~Barcza, \"O. Legeza, K.~H. Marti, and M.~Reiher.
\newblock \href{https://doi.org/10.1103/PhysRevA.83.012508}{Quantum-information
  analysis of electronic states of different molecular structures}.
\newblock {\em Phys. Rev. A}, 83:\penalty0 012508, 2011.

\bibitem{bebe-maca-2014}
M.~Bebendorf and C.~Kuske.
\newblock \href{https://doi.org/10.1007/s10915-014-9822-4}{Separation of
  variables for function generated high-order tensors}.
\newblock {\em Journal of Scientific Computing}, 61\penalty0 (1):\penalty0
  145--165, 2014.

\bibitem{botev2008efficient}
Zdravko~I Botev and Dirk~P Kroese.
\newblock \href{https://doi.org/10.1007/s11009-008-9073-7}{An efficient
  algorithm for rare-event probability estimation, combinatorial optimization,
  and counting}.
\newblock {\em Methodology and Computing in Applied Probability}, 10\penalty0
  (4):\penalty0 471--505, 2008.

\bibitem{byrne-ode-1975}
G.~D. Byrne and A.~C. Hindmarsh.
\newblock \href{https://doi.org/10.1145/355626.355636}{A polyalgorithm for the
  numerical solution of ordinary differential equations}.
\newblock {\em ACM Trans. Math. Softw.}, 1\penalty0 (1):\penalty0 71--96, 1975.

\bibitem{Cao-FSP-2016}
Youfang Cao, Anna Terebus, and Jie Liang.
\newblock \href{https://doi.org/10.1007/s11538-016-0149-1}{State space
  truncation with quantified errors for accurate solutions to discrete chemical
  master equation}.
\newblock {\em Bulletin of Mathematical Biology}, 78\penalty0 (4):\penalty0
  617--661, 2016.

\bibitem{chen-stoch-sir-2005}
Wei-Yin Chen and Sankar Bokka.
\newblock \href{https://doi.org/10.1016/j.jtbi.2004.11.033}{Stochastic modeling
  of nonlinear epidemiology}.
\newblock {\em Journal of Theoretical Biology}, 234\penalty0 (4):\penalty0
  455--470, 2005.

\bibitem{desilva-2008}
V.~de~Silva and L.-H. Lim.
\newblock \href{https://doi.org/10.1137/06066518x}{Tensor rank and the
  ill-posedness of the best low-rank approximation problem}.
\newblock {\em SIAM J. Matrix Anal. Appl.}, 30\penalty0 (3):\penalty0
  1084--1127, 2008.

\bibitem{Dinh-QTT-CME-2020}
Trang Dinh and Roger~B Sidje.
\newblock \href{https://doi.org/10.1088/1478-3975/aba1d2}{An adaptive solution
  to the chemical master equation using quantized tensor trains with sliding
  windows}.
\newblock {\em Physical Biology}, 17\penalty0 (6):\penalty0 065014, 2020.

\bibitem{dkh-cme-2014}
S.~Dolgov and B.~Khoromskij.
\newblock \href{https://doi.org/10.1002/nla.1942}{Simultaneous state-time
  approximation of the chemical master equation using tensor product formats}.
\newblock {\em Numer. Linear Algebra Appl.}, 22\penalty0 (2):\penalty0
  197--219, 2015.

\bibitem{ds-parcross-2020}
S.~Dolgov and D.~Savostyanov.
\newblock \href{https://doi.org/10.1016/j.cpc.2019.106869}{Parallel cross
  interpolation for high--precision calculation of high--dimensional
  integrals}.
\newblock {\em Comp. Phys. Comm.}, 246:\penalty0 106869, 2020.

\bibitem{dc-ttgmres-2013}
S.~V. Dolgov.
\newblock \href{https://doi.org/10.1515/rnam-2013-0009}{{TT-GMRES:} solution to
  a linear system in the structured tensor format}.
\newblock {\em Russ. J. Numer. Anal. Math. Model.}, 28\penalty0 (2):\penalty0
  149--172, 2013.

\bibitem{d-tamen-2018}
S.~V. Dolgov.
\newblock \href{https://doi.org/10.1515/cmam-2018-0023}{A tensor decomposition
  algorithm for large {ODE}s with conservation laws}.
\newblock {\em Computational Methods in Applied Mathematics}, 19:\penalty0
  23--38, 2019.

\bibitem{dkos-eigb-2014}
S.~V. Dolgov, B.~N. Khoromskij, I.~V. Oseledets, and D.~V. Savostyanov.
\newblock \href{https://doi.org/10.1016/j.cpc.2013.12.017}{Computation of
  extreme eigenvalues in higher dimensions using block tensor train format}.
\newblock {\em Computer Phys. Comm.}, 185\penalty0 (4):\penalty0 1207--1216,
  2014.

\bibitem{dks-ttfft-2012}
S.~V. Dolgov, B.~N. Khoromskij, and D.~V. Savostyanov.
\newblock \href{https://doi.org/10.1007/s00041-012-9227-4}{Superfast {Fourier}
  transform using {QTT} approximation}.
\newblock {\em J. Fourier Anal. Appl.}, 18\penalty0 (5):\penalty0 915--953,
  2012.

\bibitem{ds-amen-2014}
S.~V. Dolgov and D.~V. Savostyanov.
\newblock \href{https://doi.org/10.1137/140953289}{Alternating minimal energy
  methods for linear systems in higher dimensions}.
\newblock {\em SIAM J. Sci. Comput.}, 36\penalty0 (5):\penalty0 A2248--A2271,
  2014.

\bibitem{ds-dmrgamen-2015}
S.~V. Dolgov and D.~V. Savostyanov.
\newblock \href{https://doi.org/10.1007/978-3-319-10705-9_33}{Corrected
  one-site density matrix renormalization group and alternating minimal energy
  algorithm}.
\newblock In {\em Numerical Mathematics and Advanced Applications --- {ENUMATH}
  2013}, volume 103, 2015, pages 335--343.

\bibitem{fannes-mps-1992}
M.~Fannes, B.~Nachtergaele, and R.F. Werner.
\newblock \href{https://doi.org/10.1007/BF02099178}{Finitely correlated states
  on quantum spin chains}.
\newblock {\em Comm. Math. Phys.}, 144\penalty0 (3):\penalty0 443--490, 1992.

\bibitem{Schuette-CME-CO-2016}
Patrick Gel{\ss}, Sebastian Matera, and Christof Sch{\"u}tte.
\newblock \href{https://doi.org/10.1016/j.jcp.2016.03.025}{Solving the master
  equation without kinetic monte carlo: Tensor train approximations for a co
  oxidation model}.
\newblock {\em Journal of Computational Physics}, 314:\penalty0 489--502, 2016.

\bibitem{Gillespie77}
Daniel~T. Gillespie.
\newblock \href{https://doi.org/10.1021/j100540a008}{Exact stochastic
  simulation of coupled chemical reactions}.
\newblock {\em The Journal of Physical Chemistry}, 81\penalty0 (25):\penalty0
  2340--2361, 1977.

\bibitem{Gillespie2001}
Daniel~T. Gillespie.
\newblock \href{https://doi.org/10.1063/1.1378322}{Approximate accelerated
  stochastic simulation of chemically reacting systems}.
\newblock {\em The Journal of Chemical Physics}, 115\penalty0 (4):\penalty0
  1716--1733, 2001.

\bibitem{gleeson-degree-2011}
James~P. Gleeson.
\newblock \href{https://doi.org/10.1103/PhysRevLett.107.068701}{High-accuracy
  approximation of binary-state dynamics on networks}.
\newblock {\em Phys. Rev. Lett.}, 107:\penalty0 068701, 2011.

\bibitem{grasedyck-par-cross-2015}
L.~Grasedyck, R.~Kriemann, C.~L{\"o}bbert, A.~N{\"a}gel, G.~Wittum, and
  K.~Xylouris.
\newblock \href{https://doi.org/10.1007/s00791-015-0247-x}{Parallel tensor
  sampling in the hierarchical {Tucker} format}.
\newblock {\em Computing and Visualization in Science}, 17\penalty0
  (2):\penalty0 67--78, 2015.

\bibitem{griebel2021analysis}
Michael Griebel and Helmut Harbrecht.
\newblock \href{https://doi.org/10.1007/s10208-021-09544-6}{Analysis of tensor
  approximation schemes for continuous functions}.
\newblock {\em Foundations of Computational Mathematics}, pages 1--22, 2021.

\bibitem{Khammash-NN-CME-2021}
Ankit Gupta, Christoph Schwab, and Mustafa Khammash.
\newblock \href{https://doi.org/10.1371/journal.pcbi.1009623}{{DeepCME}: A deep
  learning framework for computing solution statistics of the chemical master
  equation}.
\newblock {\em PLOS Computational Biology}, 17\penalty0 (12):\penalty0 1--23,
  2021.

\bibitem{hackbusch-2012}
W.~Hackbusch.
\newblock {\em Tensor Spaces And Numerical Tensor Calculus}.
\newblock Springer--Verlag, Berlin, 2012.
\newblock ISBN 978-3642280269.

\bibitem{hk-ht-2009}
W.~Hackbusch and S.~{K\"uhn}.
\newblock \href{https://doi.org/10.1007/s00041-009-9094-9}{A new scheme for the
  tensor representation}.
\newblock {\em J. Fourier Anal. Appl.}, 15\penalty0 (5):\penalty0 706--722,
  2009.

\bibitem{hegland-cme-2007}
M.~Hegland, C.~Burden, L.~Santoso, S.~MacNamara, and H.~Booth.
\newblock \href{https://doi.org/10.1016/j.cam.2006.02.053}{A solver for the
  stochastic master equation applied to gene regulatory networks}.
\newblock {\em Journal of Computational and Applied Mathematics}, 205\penalty0
  (2):\penalty0 708 -- 724, 2007.

\bibitem{hegland-cme-2011}
M.~Hegland and J.~Garcke.
\newblock \href{https://doi.org/10.21914/anziamj.v52i0.3895}{On the numerical
  solution of the chemical master equation with sums of rank one tensors}.
\newblock {\em ANZIAM}, 52:\penalty0 C628--C643, 2011.

\bibitem{hemberg-perfect-sampling-2007}
M.~Hemberg and M.~Barahona.
\newblock \href{https://doi.org/10.1529/biophysj.106.099390}{Perfect sampling
  of the master equation for gene regulatory networks}.
\newblock {\em Biophysical journal}, 93\penalty0 (2):\penalty0 401--410, 2007.

\bibitem{hitchcock-sum-1927}
F.~L. Hitchcock.
\newblock The expression of a tensor or a polyadic as a sum of products.
\newblock {\em J. Math. Phys}, 6\penalty0 (1):\penalty0 164--189, 1927.

\bibitem{holtz-ALS-DMRG-2012}
S.~Holtz, T.~Rohwedder, and R.~Schneider.
\newblock \href{https://doi.org/10.1137/100818893}{The alternating linear
  scheme for tensor optimization in the tensor train format}.
\newblock {\em SIAM J. Sci. Comput.}, 34\penalty0 (2):\penalty0 A683--A713,
  2012.

\bibitem{Ion-TT-CME-2021}
Ion~Gabriel Ion, Christian Wildner, Dimitrios Loukrezis, Heinz Koeppl, and
  Herbert De~Gersem.
\newblock \href{https://doi.org/10.1063/5.0045521}{Tensor-train approximation
  of the chemical master equation and its application for parameter inference}.
\newblock {\em The Journal of Chemical Physics}, 155\penalty0 (3):\penalty0
  034102, 2021.

\bibitem{jahnke-wavelet-cme-2010}
T.~Jahnke.
\newblock \href{https://doi.org/10.1137/080742324}{An adaptive wavelet method
  for the chemical master equation}.
\newblock {\em SIAM J. Sci. Comput.}, 31\penalty0 (6):\penalty0 4373, 2010.

\bibitem{jahnke-cme-2008}
T.~Jahnke and W.~Huisinga.
\newblock \href{https://doi.org/10.1007/s11538-008-9346-x}{A dynamical low-rank
  approach to the chemical master equation}.
\newblock {\em Bulletin of Mathematical Biology}, 70:\penalty0 2283--2302,
  2008.

\bibitem{kva-anidiff-2013}
V.~Kazeev, O.~Reichmann, and Ch. Schwab.
\newblock \href{https://doi.org/10.1016/j.laa.2013.01.009}{Low-rank tensor
  structure of linear diffusion operators in the {TT} and {QTT} formats}.
\newblock {\em Linear Algebra and its Applications}, 438\penalty0
  (11):\penalty0 4204--4221, 2013.

\bibitem{khkaz-lap-2012}
V.~A. Kazeev and B.~N. Khoromskij.
\newblock \href{https://doi.org/10.1137/100820479}{Low-rank explicit {QTT}
  representation of the {L}aplace operator and its inverse}.
\newblock {\em SIAM J. Matrix Anal. Appl.}, 33\penalty0 (3):\penalty0 742--758,
  2012.

\bibitem{kkns-cme-2014}
Vladimir Kazeev, Mustafa Khammash, Michael Nip, and Christoph Schwab.
\newblock \href{https://doi.org/10.1371/journal.pcbi.1003359}{Direct solution
  of the {C}hemical {M}aster {E}quation using {Q}uantized {T}ensor {T}rains}.
\newblock {\em PLOS Computational Biology}, 10\penalty0 (3):\penalty0 e100359,
  2014.

\bibitem{kkns-cme-theory-2015}
Vladimir Kazeev and Christoph Schwab.
\newblock \href{https://doi.org/10.1137/130927218}{Tensor approximation of
  stationary distributions of chemical reaction networks}.
\newblock {\em SIAM Journal on Matrix Analysis and Applications}, 36\penalty0
  (3):\penalty0 1221--1247, 2015.

\bibitem{keeling-meanfield-1999}
M.~J. Keeling.
\newblock \href{https://doi.org/10.1098/rspb.1999.0716}{The effects of local
  spatial structure on epidemiological invasions}.
\newblock {\em Proc Biol Sci}, 266\penalty0 (1421):\penalty0 859--867, 1999.

\bibitem{KemenySnell-MC-1976}
John~G. Kemeny and J.~Laurie Snell.
\newblock {\em Finite Markov Chains}.
\newblock Springer, 1976.

\bibitem{kmk-sir-1927}
William~Ogilvy Kermack and Anderson~Gray McKendrick.
\newblock \href{https://doi.org/10.1098/rspa.1927.0118}{A contribution to the
  mathematical theory of epidemics}.
\newblock {\em Proceedings of the Royal Society London A}, 115\penalty0
  (772):\penalty0 700--721, 1927.

\bibitem{bokh-surv-2015}
B.~N. Khoromskij.
\newblock \href{https://doi.org/10.1051/proc/201448001}{Tensor numerical
  methods for multidimensional {PDE}s: theoretical analysis and initial
  applications}.
\newblock {\em ESAIM: Proc.}, 48:\penalty0 1--28, 2015.

\bibitem{klumper-mps-1993}
A.~Kl\"umper, A.~Schadschneider, and J.~Zittartz.
\newblock \href{https://doi.org/10.1209/0295-5075/24/4/010}{Matrix product
  ground states for one-dimensional spin-1 quantum antiferromagnets}.
\newblock {\em Europhys. Lett.}, 24\penalty0 (4):\penalty0 293--297, 1993.

\bibitem{kolda-review-2009}
T.~G. Kolda and B.~W. Bader.
\newblock \href{https://doi.org/10.1137/07070111X}{Tensor decompositions and
  applications}.
\newblock {\em SIAM Rev.}, 51\penalty0 (3):\penalty0 455--500, 2009.

\bibitem{Schuette-RBF-CME-2015}
Ivan Kryven, Susanna R{\"o}blitz, and Christof Sch{\"u}tte.
\newblock \href{https://doi.org/10.1186/s12918-015-0210-y}{Solution of the
  chemical master equation by radial basis functions approximation with
  interface tracking}.
\newblock {\em BMC Systems Biology}, 9\penalty0 (1):\penalty0 67, 2015.

\bibitem{Yates-MLCME-2016}
Christopher Lester, Ruth~E. Baker, Michael~B. Giles, and Christian~A. Yates.
\newblock \href{https://doi.org/10.1007/s11538-016-0178-9}{Extending the
  multi-level method for the simulation of stochastic biological systems}.
\newblock {\em Bulletin of Mathematical Biology}, 78\penalty0 (8):\penalty0
  1640--1677, 2016.

\bibitem{lindquist-degree-2011}
Jennifer Lindquist, Junling Ma, P.~van~den Driessche, and Frederick~H.
  Willeboordse.
\newblock \href{https://doi.org/10.1007/s00285-010-0331-2}{Effective degree
  network disease models}.
\newblock {\em J Math Biol}, 62:\penalty0 143--164, 2011.

\bibitem{miller-edge-2012}
Joel~C. Miller, Anja~C. Slim, and Erik~M. Volz.
\newblock \href{https://doi.org/10.1098/rsif.2011.0403}{Edge-based
  compartmental modelling for infectious disease spread}.
\newblock {\em J. R. Soc. Interface}, 9:\penalty0 890–906, 2012.

\bibitem{munsky-fsp-2006}
B.~Munsky and M.~Khammash.
\newblock \href{https://doi.org/10.1063/1.2145882}{The finite state projection
  algorithm for the solution of the chemical master equation}.
\newblock {\em The Journal of chemical physics}, 124:\penalty0 044104, 2006.

\bibitem{osel-tt-2011}
I.~V. Oseledets.
\newblock \href{https://doi.org/10.1137/090752286}{Tensor-train decomposition}.
\newblock {\em SIAM J. Sci. Comput.}, 33\penalty0 (5):\penalty0 2295--2317,
  2011.

\bibitem{ost-latensor-2009}
I.~V. Oseledets, D.~V. Savostyanov, and E.~E. Tyrtyshnikov.
\newblock \href{https://doi.org/10.1007/s00607-009-0047-6}{Linear algebra for
  tensor problems}.
\newblock {\em Computing}, 85\penalty0 (3):\penalty0 169--188, 2009.

\bibitem{ot-ttcross-2010}
I.~V. Oseledets and E.~E. Tyrtyshnikov.
\newblock \href{https://doi.org/10.1016/j.laa.2009.07.024}{{TT-cross}
  approximation for multidimensional arrays}.
\newblock {\em Linear Algebra Appl.}, 432\penalty0 (1):\penalty0 70--88, 2010.

\bibitem{peherstorfer2018multifidelity}
Benjamin Peherstorfer, Boris Kramer, and Karen Willcox.
\newblock \href{https://doi.org/10.1137/17M1122992}{Multifidelity
  preconditioning of the cross-entropy method for rare event simulation and
  failure probability estimation}.
\newblock {\em SIAM/ASA Journal on Uncertainty Quantification}, 6\penalty0
  (2):\penalty0 737--761, 2018.

\bibitem{rno-gpueig-2019}
Maxim Rakhuba, Alexander Novikov, and Ivan Oseledets.
\newblock \href{https://doi.org/10.1016/j.jcp.2019.07.003}{Low-rank
  {Riemannian} eigensolver for high-dimensional {Hamiltonians}}.
\newblock {\em J. Comput. Phys.}, 396\penalty0 (1):\penalty0 718--737, 2019.

\bibitem{rand-meanfield-1999}
D.~A. Rand.
\newblock Correlation equations and pair approximations for spatial ecologies.
\newblock In {\em Advanced Ecological Theory: Principles and Applications},
  Blackwell Science, Oxford, 1999, chapter~4, pages 100--142.

\bibitem{RogersPitman-MC-1981}
L.~C.~G. Rogers and J.~W. Pitman.
\newblock \href{https://doi.org/10.1214/aop/1176994363}{Markov functions}.
\newblock {\em Ann. Probab.}, 9\penalty0 (4):\penalty0 573--582, 1981.

\bibitem{sav-rank1-2012}
D.~V. Savostyanov.
\newblock \href{https://doi.org/10.1016/j.laa.2011.11.008}{{QTT}-rank-one
  vectors with {QTT}-rank-one and full-rank {Fourier} images}.
\newblock {\em Linear Algebra Appl.}, 436\penalty0 (9):\penalty0 3215--3224,
  2012.

\bibitem{sav-qott-2014}
D.~V. Savostyanov.
\newblock \href{https://doi.org/10.1016/j.laa.2014.06.006}{Quasioptimality of
  maximum--volume cross interpolation of tensors}.
\newblock {\em Linear Algebra Appl.}, 458:\penalty0 217--244, 2014.

\bibitem{so-dmrgi-2011proc}
D.~V. Savostyanov and I.~V. Oseledets.
\newblock \href{https://doi.org/10.1109/nDS.2011.6076873}{Fast adaptive
  interpolation of multi-dimensional arrays in tensor train format}.
\newblock In {\em Proceedings of 7th International Workshop on Multidimensional
  Systems (nDS)}, IEEE, 2011.

\bibitem{schollwock-2011}
U.~Schollw\"ock.
\newblock \href{https://doi.org/10.1016/j.aop.2010.09.012}{The density-matrix
  renormalization group in the age of matrix product states}.
\newblock {\em Annals of Physics}, 326\penalty0 (1):\penalty0 96--192, 2011.

\bibitem{sgldcj-pTDVP-2020}
P.~Secular, N.~Gourianov, M.~Lubasch, S.~Dolgov, S.~R. Clark, and D.~Jaksch.
\newblock \href{https://doi.org/10.1103/PhysRevB.101.235123}{Parallel
  time-dependent variational principle algorithm for matrix product states}.
\newblock {\em Phys. Rev. B}, 101:\penalty0 235123, 2020.

\bibitem{solomonik-cyclops-2014}
Edgar Solomonik, Devin Matthews, Jeff~R. Hammond, John~F. Stanton, and James
  Demmel.
\newblock \href{https://doi.org/10.1016/j.jpdc.2014.06.002}{A massively
  parallel tensor contraction framework for coupled-cluster computations}.
\newblock {\em Journal of Parallel and Distributed Computing}, 74\penalty0
  (12):\penalty0 3176 -- 3190, 2014.

\bibitem{Grima-CME-NN-2022}
Augustinas Sukys, Kaan {\"O}cal, and Ramon Grima.
\newblock \href{https://doi.org/10.1101/2022.04.26.489548}{Approximating
  solutions of the chemical master equation using neural networks}.
\newblock {\em bioRxiv}, 2022.

\bibitem{kiss-degree-2012}
Michael Taylor, Timothy~J. Taylor, and Istvan~Z. Kiss.
\newblock \href{https://doi.org/10.1103/PhysRevE.85.016103}{Epidemic threshold
  and control in a dynamic network}.
\newblock {\em Phys. Rev. E}, 85:\penalty0 016103, 2012.

\bibitem{trefethen-spectral-2000}
L.~N. Trefethen.
\newblock {\em Spectral methods in MATLAB}.
\newblock SIAM, Philadelphia, 2000.

\bibitem{vankampen-stochastic-1981}
N.~G. van Kampen.
\newblock {\em Stochastic processes in physics and chemistry}.
\newblock North Holland, Amsterdam, 1981.

\bibitem{Sidje-TT-CME-2017}
Huy~D. Vo and Roger~B. Sidje.
\newblock \href{https://doi.org/10.1063/1.4994917}{An adaptive solution to the
  chemical master equation using tensors}.
\newblock {\em The Journal of Chemical Physics}, 147\penalty0 (4):\penalty0
  044102, 2017.

\bibitem{wagner2020multilevel}
Fabian Wagner, Jonas Latz, Iason Papaioannou, and Elisabeth Ullmann.
\newblock \href{https://doi.org/10.1137/19M1289601}{Multilevel sequential
  importance sampling for rare event estimation}.
\newblock {\em SIAM Journal on Scientific Computing}, 42\penalty0 (4):\penalty0
  A2062--A2087, 2020.

\bibitem{white-dmrg-1993}
Steven~R. White.
\newblock \href{https://doi.org/10.1103/PhysRevB.48.10345}{Density-matrix
  algorithms for quantum renormalization groups}.
\newblock {\em Phys. Rev. B}, 48\penalty0 (14):\penalty0 10345--10356, 1993.

\bibitem{youssef-network-sir-2011}
Mina Youssef and Caterina Scoglio.
\newblock \href{https://doi.org/10.1016/j.jtbi.2011.05.029}{An
  individual--based approach to {SIR} epidemics in contact networks}.
\newblock {\em Journal of Theoretical Biology}, 283\penalty0 (1):\penalty0
  136--144, 2011.

\end{thebibliography}

\end{document}